\documentclass[11pt,preprint]{elsarticle}

\usepackage{fancyhdr}
\usepackage{fullpage}
\usepackage{amsmath}
\usepackage{amsbsy}
\usepackage{amssymb}
\usepackage{amscd}
\usepackage{amsfonts}
\usepackage{supertabular}
\usepackage{graphics}
\usepackage{verbatim}
\usepackage{epsfig}
\usepackage{xspace}
\usepackage{euscript}
\usepackage{alltt}
\usepackage{boxedminipage}
\usepackage{float}
\usepackage[colorlinks]{hyperref}
\usepackage{color}
\usepackage[all]{xy}
\usepackage{t1enc}
\usepackage{times,exscale}
\usepackage{graphicx,calc}
\usepackage{subfig}
\usepackage[ruled,vlined]{algorithm2e}
\usepackage{epstopdf}

\def\bR{{\mathbf{R}}}
\def\bx{{\mathbf{x}}}
\def\R{{\mathbb{R}}}

\newcommand{\norm}[1]{\| #1 \|}

\begin{document}

\begin{frontmatter}

\title{Higher-order finite-difference formulation of periodic Orbital-free Density Functional Theory}
\author[gatech]{Swarnava Ghosh}
\author[gatech]{Phanish Suryanarayana\corref{cor}}
\address[gatech]{College of Engineering, Georgia Institute of Technology, GA 30332, USA}
\cortext[cor]{Corresponding Author (\it phanish.suryanarayana@ce.gatech.edu) }

\begin{abstract}
We present a real-space formulation and higher-order finite-difference implementation of periodic Orbital-free Density Functional Theory (OF-DFT). Specifically, utilizing a local reformulation of the electrostatic and kernel terms, we develop a generalized framework for performing OF-DFT simulations with different variants of the electronic kinetic energy. In particular, we propose a self-consistent field (SCF) type fixed-point method for calculations involving linear-response kinetic energy functionals. In this framework, evaluation of both the electronic ground-state as well as forces on the nuclei are amenable to computations that scale linearly with the number of atoms. We develop a parallel implementation of this formulation using the finite-difference discretization. We demonstrate that higher-order finite-differences can achieve relatively large convergence rates with respect to mesh-size in both the energies and forces. Additionally, we establish that the fixed-point iteration converges rapidly, and that it can be further accelerated using extrapolation techniques like Anderson's mixing. We validate the accuracy of the results by comparing the energies and forces with plane-wave methods for selected examples, including the vacancy formation energy in Aluminum. Overall, the suitability of the proposed formulation for scalable high performance computing makes it an attractive choice for large-scale OF-DFT calculations consisting of thousands of atoms. 
\end{abstract}


\end{frontmatter}

\section{Introduction}
Kohn-Sham Density Functional Theory (DFT) \cite{Hohenberg, Kohn1965} has a relatively high accuracy/cost ratio, which makes it a popular electronic structure method for predicting material properties and behavior. In DFT, the system of interacting electrons is replaced with a system of non-interacting electrons moving in an effective potential \cite{Parr1989,NumAnalysis2003}. The electronic ground-state in DFT is typically determined by solving for the Kohn-Sham orbitals, the number of which is commensurate with the size of the system, i.e. number of electrons \cite{NumAnalysis2003,Martin2004}. Since these orbitals need to be orthonormal, the overall solution procedure scales cubically with the number of atoms \cite{NumAnalysis2003,Martin2004}. In order to overcome this restrictive scaling, significant research has focused on the development of linear-scaling methods \cite{Goedecker, Bowler2012}. Nearly all of these approaches, in one form or the other, employ the decay of the density matrix \cite{benzi2013decay} in conjunction with truncation to achieve linear-scaling \cite{Goedecker, Bowler2012}. However, an efficient linear-scaling algorithm for metallic systems at low temperatures still remains an open problem \cite{Bowler2012,Cances2008}.

Orbital-free DFT (OF-DFT) represents a simplified version of DFT, wherein the electronic kinetic energy is modeled using a functional of the electron density \cite{WangBook2002}. Commonly used kinetic energy functionals include the Thomas-Fermi-von Weizsacker (TFW) \cite{Thomas1927,Fermi1927,Weizsacker1935}, Wang-Teter (WT) \cite{Teter1992}, and Wang, Govind \& Carter (WGC) \cite{Wang1998,Wang1999} variants. Amongst these, the WT and WGC functionals are designed so as to match the linear-response of a homogeneous electron gas \cite{WangBook2002}. Previous studies have shown that OF-DFT is able to provide an  accurate description of systems whose electronic structure resembles a free-electron gas, e.g. Aluminum and Magnesium \cite{carling2003orbital,ho2007energetics,huang2008transferable}. There have been recent efforts to extend the applicability of OF-DFT to covalently bonded materials \cite{zhou2005improving} as well as molecular systems \cite{xia2012can}. In essence, OF-DFT can be viewed as a `single-orbital' version of DFT, wherein the cubic-scaling bottleneck arising from orthogonalization is no longer applicable. In addition, OF-DFT possesses an extremely favorable scaling with respect to temperature compared to DFT \cite{sjostrom2014fast,karasiev2014finite}. Overall, OF-DFT has the potential to enable electronic structure calculations for system sizes that are intractable for DFT. 

The plane-wave basis is attractive for performing OF-DFT calculations \cite{Teter1992,Ho2008,Hung2010} because of the spectral convergence with increasing basis size and the efficient evaluation of convolutions using the Fast Fourier Transform (FFT) \cite{Cooley1965}. However, developing implementations which can efficiently utilize modern large-scale, distributed-memory computer architectures is particularly challenging. Further, evaluation of the electrostatic terms within the plane-wave basis typically scales quadratically with the number of atoms \cite{hung2009accurate}. In view of this, recent efforts have been directed towards developing real-space approaches for OF-DFT, including finite-differences \cite{Suryanarayana2014524} and finite-elements \cite{Gavini2007,Motamarri2012}. Amongst these, the finite-element method provides the flexibility of an adaptive discretization. This attribute has been employed to perform all-electron calculations \cite{Gavini2007,Motamarri2012} and to develop a coarse-grained formulation of OF-DFT for studying crystal defects \cite{GaviniQC2007}. However, higher-order finite-differences---shown to be extremely efficient in non-periodic OF-DFT with the TFW kinetic energy functional \cite{Suryanarayana2014524}---remain unexplored in the context of periodic OF-DFT simulations, particularly when linear-response kinetic energy functionals like WT and WGC are employed. 

The electronic ground state in OF-DFT can be expressed as the solution of a non-linear, constrained minimization problem \cite{Garcia2007,Lieb1981,Blanc2005,Cances2012,Gavini2007,Suryanarayana2014524}. The approaches which have previously been employed to solve this problem include variants of conjugate-gradient \cite{Ho2008,Gavini2007,Jiang2004,Suryanarayana2014524} and Newton \cite{Ho2008,Garcia2007,Motamarri2012} methods. In these approaches, the techniques used to enforce the constraints include Lagrange multipliers \cite{hung2012preconditioners,Motamarri2012}, the penalty method \cite{Gavini2007} and the Augmented-Lagrangian method \cite{Suryanarayana2014524}. In this work, we present a local real-space formulation and implementation of periodic OF-DFT. In particular, we develop a fixed-point iteration with respect to the kernel potential for simulations involving linear-response kinetic energy functionals. We develop a parallel implementation of the formulation using higher-order finite-differences. We demonstrate the robustness, efficiency and accuracy of the proposed approach through selected examples, the results of which are compared against existing plane-wave methods. 

The remainder of this paper is organized as follows. We introduce OF-DFT in Section \ref{Sec:OFDFT} and discuss its real-space formulation in Section \ref{Sec:Formulation}. Subsequently, we describe the numerical implementation in Section \ref{Section:NumericalImplementation}, and validate it through examples in Section \ref{Section:Examples}. Finally, we conclude in Section \ref{Section:Conclusions}.

\section{Orbital-free Density Functional Theory} \label{Sec:OFDFT}
Consider a charge neutral system of $M_a$ atoms and $N_e$ electrons in a cuboidal domain $\Omega$ under periodic boundary conditions. Let $\bR = \{\bR_1, \bR_2, \ldots, \bR_{M_a} \}$ denote the positions of the nuclei with charges $\mathbf{Z} = \{Z_1, Z_2, \ldots, Z_{M_a} \}$ respectively. The energy of this system as described by OF-DFT is \cite{Parr1989}
\begin{equation} \label{Eqn:Energy:OFDFT}
\mathcal{E} (u, \bR) = T_s(u) + E_{xc}(u) + E_{H}(u) + E_{ext}(u,\bR) + E_{zz}(\bR) \,,
\end{equation}
where $u = \sqrt{\rho}$, $\rho$ being the electron density. Introducing the parameters $\lambda$ and $\mu \in \{0,1\}$ so that different variants of the electronic kinetic energy $T_s(u)$ can be encompassed within a single expression, we can write
\begin{eqnarray} \label{Eqn:KineticEnergyComponents}
T_s(u) = T_{TF}(u) + \lambda T_{vW}(u) + \mu T_{LR}(u) \,,
\end{eqnarray}
where $T_{TF}(u)$ is the Thomas-Fermi energy \cite{Thomas1927,Fermi1927}, $T_{vW}(u)$ is the von Weizsacker \cite{Weizsacker1935} term and $T_{LR}(u)$ is a non-local kernel energy incorporated to make the kinetic energy satisfy the linear-response of a homogeneous electron gas \cite{WangBook2002}. They can be represented as 
\begin{eqnarray}
T_{TF}(u) & = & C_F \int_{\Omega} u^{10/3} (\bx) \, \mathrm{d\bx} \,, \label{Eqn:TF:KineticEnergy}\\
T_{vW}(u) & = & \frac{1}{2}\int_{\Omega} |\nabla u(\bx)|^2 \, \mathrm{d\bx} \,, \label{Eqn:vW:KineticEnergy} \\
T_{LR}(u) & = & C_F \int_{\R^3} \int_{\Omega} u^{2\alpha}(\bx) \, K(|\bx-\bx'|,\rho(\bx),\rho(\bx')) \, u^{2\beta}(\bx') \, \mathrm{d\bx} \, \mathrm{d\bx'} \,, \label{Eqn:LR:KineticEnergy}
\end{eqnarray}
where $\alpha$ and $\beta$ are parameters, and the constant $C_F = \frac{3}{10} (3\pi^2)^\frac{2}{3}$. On the one hand, the Thomas-Fermi-von Weizsacker (TFW) family of functionals with the adjustable parameter $\lambda$ is obtained by setting $\mu=0$ \cite{Parr1989}. On the other hand, kinetic energy functionals which satisfy the Lindhard susceptibility function are obtained by setting $\mu=\lambda=1$ with appropriate choices of $\alpha$, $\beta$ and the kernel $K(|\bx-\bx'|,\rho(\bx),\rho(\bx'))$ \cite{WangBook2002}. In particular, the Wang \& Teter (WT) functional \cite{Teter1992} utilizes a density independent kernel, whereas the Wang, Govind \& Carter (WGC) functional \cite{Wang1998,Wang1999} employs a density dependent kernel. It is common to perform a Taylor series expansion of the density dependent kernel $K(|\bx-\bx'|,\rho(\bx),\rho(\bx'))$ about the average electron density $\bar{\rho}$ \cite{Wang1999}. On doing so, we arrive at 
\begin{eqnarray}\label{Eqn:LR:KineticEnergy:Linearized}
T_{LR}(u) = C_F \sum_{m=0}^{L} \sum_{n=0}^{L} \sum_{p=0}^m \sum_{q=0}^n  C_{mnpq} \int_{\R^3} \int_{\Omega} u^{2(m-p+\alpha)}(\bx)K_{mn}(|\bx-\bx'|)u^{2(n-q+\beta)}(\bx') \,\mathrm{d\bx} \, \mathrm{d\bx'} \,,
\end{eqnarray} 
where $L$ is the order of the expansion, the coefficients
\begin{equation}
C_{mnpq} = \frac{(-1)^{p+q}}{m!\,n!}\binom{m}{p} \binom{n}{q} \bar{\rho}^{p+q-m-n} \,,
\end{equation}
and the kernels
\begin{equation}
K_{mn}(|\bx-\bx'|) = \bar{\rho}^{m+n} \left( \frac{\partial^{m+n}K(|\bx-\bx'|,\rho(\bx),\rho(\bx'))}{\partial \rho^m(\bx) \partial \rho^n(\bx')} \right) \bigg|_{\rho=\bar{\rho}} .
\end{equation}

The second term in Eqn. \ref{Eqn:Energy:OFDFT} is referred to as the exchange-correlation energy. It is generally modeled in OF-DFT using the local density approximation (LDA) \cite{Kohn1965}:
\begin{equation}
E_{xc} (u) = \int_{\Omega} \varepsilon_{xc} (u(\bx)) u^2(\bx) \, \mathrm{d \bx} \,,
\end{equation}
where $\varepsilon_{xc} (u) = \varepsilon_x (u) + \varepsilon_c (u)$ is the sum of the exchange and correlation per particle of a uniform electron gas of density $\rho = u^2 $. Employing the Perdew-Zunger \cite{PhysRevB.23.5048} parameterization of the correlation energy calculated by Ceperley-Alder \cite{Ceperley1980}, the exchange and correlation functionals can be represented as
\begin{eqnarray}
\varepsilon_{x}(u) & = & -\frac{3}{4}\left(\frac{3}{\pi}\right)^{1/3}u^{2/3} \,, \\
\varepsilon_{c}(u) & = & 
\begin{cases}
\frac{\gamma_1}{1+\beta_1\sqrt{r_s}+\beta_2{r_s}}\,\,\,\,\,\,\,\,\,\,\,\
r_s\geq 1\\
A_1\log{r_s}+B_1+C_1r_s\log{r_s}+D_1r_s\,\,\,\, r_s<1
\end{cases}
\end{eqnarray}
where $r_s=(\frac{3}{4\pi u^2})^{1/3}$, and the constants $\gamma_1=-0.1423$, ${\beta_1}=1.0529$, ${\beta_2}=0.3334$, $A_1=0.0311$, $B_1=-0.048$, $C_1=0.002$ and $D_1=-0.0116$.

The final three terms in Eqn. \ref{Eqn:Energy:OFDFT} represent electrostatic energies \cite{Martin2004}. In periodic systems, they can be expressed as 
\begin{eqnarray}
E_{H}(u) & = & \frac{1}{2} \int_{\R^3} \int_{\Omega} \frac{u^2(\bx)u^2(\bx')}{|\bx - \bx'|} \,\mathrm{d\bx} \, \mathrm{d\bx'}\,, \label{Eqn:EH} \\
E_{ext} (u,\bR) & = & \sum_{I} \int_{\Omega} u^2(\bx) V_{I}(\bx,\bR_I)  \, \mathrm{d\bx} \,, \label{Eqn:Eext}\\
E_{zz}(\bR) & = & \frac{1}{2} \sum_{I} \sum_{\begin{subarray}{c} J_{\Omega} \\J_{\Omega} \neq I \end{subarray}} \frac{Z_{I} Z_{J_{\Omega}}}{|\bR_{I}-\bR_{J_{\Omega}}|} \,, \label{Eqn:EZZ}
\end{eqnarray}
where the summation indices $I$ and $J_{\Omega}$ run over all atoms in $\R^3$ and $\Omega$, respectively. The Hartree energy $E_{H}(u)$ is the classical interaction energy of the electron density, $V_{I}(\bx,\bR_I)$ is the potential due to the nucleus positioned at $\bR_I$, $E_{ext}(u,\bR)$ is the interaction energy between the electron density and the nuclei, and $E_{zz}(\bR)$ is the repulsion energy between the nuclei. 

The ground state of the system in OF-DFT is given by the variational problem \cite{Lieb1981, Gavini2007, Garcia2007, Cances2012,Suryanarayana2014524}
\begin{eqnarray} \label{Eqn:VariationalProblem:OFDFT}
\mathcal{E}_0 =  \inf_{\bR \in \R^{3M_a}} \inf_{u \in \mathcal{X}} \mathcal{E}(u,\bR)  \,, \quad \mathcal{X} = \left \{u: u\in X,\, u \geq 0, \, \mathcal{C}(u)=0 \right \} \,,
\end{eqnarray} 
where $X$ is some appropriate space of periodic functions and 
\begin{equation}
\mathcal{C}(u) = \int_{\Omega}{u^2(\bx)} \, \mathrm{d\bx} - N_e
\end{equation} 
represents the constraint on the total number of electrons. The inequality constraint $u \geq 0$ is to ensure that $u$ is nodeless, i.e. $u$ does not change sign. In this work, we focus on developing a local formulation and higher-order finite-difference implementation for determining the ground-state in periodic OF-DFT simulations.   


\section{Real-space formulation} \label{Sec:Formulation}
In this section, we develop a framework for periodic OF-DFT that is amenable to a linear-scaling real-space implementation. First, we present a local description of the kernel energy and potential in Section \ref{Subsec:LocalReformulationVLR}. Next, we discuss how the electrostatics can be rewritten into local form in Section \ref{Subsec:ElectrostaticReformulation}. Finally, we describe the methodology for determining the OF-DFT ground-state in Section \ref{Subsec:GroundState}. 

\subsection{Local reformulation of the kernel energy and potential} \label{Subsec:LocalReformulationVLR}
In simulations where linear-response kinetic energy functionals are employed, the kernel energy $T_{LR}(u)$ as well as the kernel potential 
\begin{eqnarray}
V_{LR} (\bx) & = & \frac{\delta T_{LR}(u)}{\delta u^2} \nonumber \\
& = & C_F \sum_{m=0}^{L} \sum_{n=0}^{L} \sum_{p=0}^m \sum_{q=0}^n  C_{mnpq} \bigg[ (m-p+\alpha)  u^{2(m-p+\alpha-1)}(\bx) \int_{\R^3} K_{mn}(|\bx-\bx'|) u^{2(n-q+\beta)}(\bx') \, \mathrm{d\bx'} \nonumber \\  
 &  + & (n-q+\beta)  u^{2(n-q+\beta-1)}(\bx) \int_{\R^3} K_{mn}(|\bx-\bx'|) u^{2(m-p+\alpha)}(\bx') \, \mathrm{d\bx'} \Bigg] \,,
\end{eqnarray}
are inherently non-local in real-space. In order to enable a linear-scaling implementation, we start by defining the potentials
\begin{eqnarray}
V_{mnq\beta} (\bx) & = & \int_{\R^3} K_{mn}(|\bx-\bx'|) u^{2(n-q+\beta)}(\bx') \, \mathrm{d\bx'}  \,, \\
V_{mnp\alpha} (\bx) & = & \int_{\R^3} K_{mn}(|\bx-\bx'|) u^{2(m-p+\alpha)}(\bx') \, \mathrm{d\bx'} \,. 
\end{eqnarray}
After approximating the kernels $K_{mn}(|\bx-\bx'|)$ in Fourier space using rational functions \cite{Choly2002}, we arrive at 
\begin{eqnarray}
V_{mnq\beta}(\bx) & = & \sum_{r=1}^R V_{mnq\beta r} (\bx) \,, \\
V_{mnp\alpha}(\bx) & = & \sum_{r=1}^R V_{mnp\alpha r}(\bx) \,,
\end{eqnarray}
where $V_{mnq\beta r}(\bx)$ and $V_{mnp\alpha r}(\bx)$ are solutions of the Helmholtz equations
\begin{eqnarray}
-\frac{1}{(2\bar{k}_F)^2}\nabla^2  V_{mnq \beta r}(\bx) + Q_{mnr} V_{mnq \beta r}(\bx) & = & P_{mnr} f_{mp\alpha}(\bx) \,, \label{Eqn:Helmholtz:beta}\\
-\frac{1}{(2\bar{k}_F)^2}\nabla^2  V_{mnp \alpha r}(\bx) + Q_{mnr} V_{mnp \alpha r}(\bx) & = & P_{mnr} f_{nq\beta}(\bx) \,, \label{Eqn:Helmholtz:alpha}
\end{eqnarray}
under periodic boundary conditions and appropriate choice of complex constants $P_{mnr}$ and $Q_{mnr}$. Above, $\bar{k}_F=(3\pi^2\bar{\rho})^{\frac{1}{3}}$ and 
\begin{eqnarray}
f_{mp\alpha}(\bx) & = &  
\begin{cases}
-\frac{1}{(2\bar{k}_F)^2} \nabla^2 u^{2(m-p+\alpha)}(\bx) &  \text{if} \,\,\, m = n = 0 \,,  \\
u^{2(m-p+\alpha)}(\bx)  & \text{otherwise} \,, 
\end{cases} \\
f_{nq\beta}(\bx) & = &  
\begin{cases}
-\frac{1}{(2\bar{k}_F)^2} \nabla^2 u^{2(n-q+\beta)}(\bx) &  \text{if} \,\,\, m = n = 0 \,,  \\
u^{2(n-q+\beta)}(\bx)  & \text{otherwise} \,.
\end{cases} 
\end{eqnarray}

Thereafter, the kernel potential $V_{LR}(\bx)$ and the corresponding kernel energy $T_{LR}(u)$ can be calculated in linear-scaling fashion using the expressions
\begin{eqnarray}
V_{LR} (\bx) & = & C_F \sum_{m=0}^{L} \sum_{n=0}^{L} \sum_{p=0}^m \sum_{q=0}^n  \sum_{r=1}^R  C_{mnpq} \bigg[ (m-p+\alpha)  u^{2(m-p+\alpha-1)}(\bx) V_{mnq\beta r}(\bx) \, \nonumber \\ & + & (n-q+\beta) u^{2(n-q+\beta-1)}(\bx) V_{mnp\alpha r}(\bx) \Bigg] \,, \label{Eqn:VLinearResponse:local} \\
T_{LR}(u) & = & \frac{1}{2}C_F \sum_{m=0}^{L} \sum_{n=0}^{L} \sum_{p=0}^m \sum_{q=0}^n  \sum_{r=1}^R C_{mnpq} \int_{\Omega} \bigg[ u^{2(m-p+\alpha)}(\bx) V_{mnq\beta r}(\bx)   +  u^{2(n-q+\beta)}(\bx) V_{mnp\alpha r}(\bx) \bigg] \, \mathrm{d\bx} \nonumber \\ 
\end{eqnarray}
where $V_{mnq\beta r}(\bx)$ and $V_{mnp\alpha r}(\bx)$ are solutions of the Helmholtz equations given in Eqns. \ref{Eqn:Helmholtz:beta} and \ref{Eqn:Helmholtz:alpha}, respectively.


\subsection{Local reformulation of the electrostatics}\label{Subsec:ElectrostaticReformulation}
The electrostatic energies in Eqns. \ref{Eqn:EH}, \ref{Eqn:Eext} and \ref{Eqn:EZZ} are non-local in real-space. Moreover, they are individually divergent in periodic systems. To overcome this, we introduce the charge density of the nuclei \cite{Pask2005,Gavini2007,Phanish2012,Suryanarayana2014524}:
\begin{equation} \label{Eqn:b:Pseudopotential}
\quad b(\bx,\bR) = \sum_{J} b_J (\bx,\bR_J) \,\,, \,\, \,\, b_J(\bx,\bR_J) = \frac{-1}{4 \pi} \nabla^2 V_J (\bx,\bR_J) \,,
\end{equation}
where $b_J(\bx,\bR_J)$ is the charge density of the $J^{th}$ nucleus, and the summation index $J$ runs over all atoms in $\R^3$. In OF-DFT calculations, it is common to remove the core electrons and replace the singular Coulomb potential with an effective potential $V_J(\bx,\bR_J)$, referred to as the pseudopotential approximation \cite{pickett1989pseudopotential}. The absence of orbitals in OF-DFT requires that the pseudopotential be local, i.e. $V_J(\bx,\bR_J)$ depends only on the distance from the nucleus. Since the pseudopotential replicates the Coulomb potential outside the core cutoff radius $r_c$, $b_J(\bx,\bR_J)$ has a compact support within a ball of radius $r_c$ centered at $\bR_J$ \cite{Pask2005,Suryanarayana2014524}. It follows that 
\begin{equation}
\int_{\R^3} b_J(\bx,\bR_J) \, \mathrm{d\bx} = Z_J \,, \quad \int_{\Omega} b(\bx,\bR) \, \mathrm{d\bx} = N_e \,.
\end{equation}

Using the above definition for the charge densities, we can rewrite the total electrostatic energy as the following variational problem
\begin{eqnarray} \label{Eqn:LocalReformulationElectrostaticEnergy}
E_H(u) + E_{ext}(u,\bR) + E_{zz}(\bR) & = & \sup_{\phi \in Y}  \bigg \{ - \frac{1}{8 \pi} \int_{\Omega} |\nabla \phi(\bx)|^2 \, \mathrm{d\bx} + \int_{\Omega}(u^2(\bx)+ b(\bx,\bR)) \phi(\bx) \, \mathrm{d\bx} \bigg \} \nonumber \\ 
& - & \frac{1}{2}\sum_{J} \int_{\Omega} b_J(\bx,\bR_J) V_J(\bx,\bR_J) \, \mathrm{d\bx}  + \mathcal{E}_c^*(\bR) \,, \label{Eqn:ElecEnergyReformulation}      
\end{eqnarray}
where $\phi(\bx)$ is the electrostatic potential, $Y$ is some appropriate space of periodic functions, the second last term accounts for the self energy of the nuclei and the last term corrects for overlapping charge densities. A detailed discussion on the nature of $\mathcal{E}_c^*(\bR)$ and its evaluation can be found in Appendix \ref{Appendix:Correct:RepulsiveEnergy}. With this reformulation of the total electrostatic energy, we arrive at the variational problem
\begin{equation}
\mathcal{E}(u,\bR) = \bigg\{ \sup_{\phi \in Y} \mathcal{F}(u,\bR,\phi) + \mu T_{LR}(u) \bigg\} \,,
\end{equation}
where the functional
\begin{eqnarray}
\mathcal{F}(u,\bR,\phi)  & = & C_F \int_{\Omega} u^{10/3} (\bx) \, \mathrm{d\bx} + \frac{\lambda}{2} \int_{\Omega} |\nabla u(\bx)|^2 \, \mathrm{d\bx} + \int_{\Omega} \varepsilon_{xc} (u(\bx)) u^2(\bx) \, \mathrm{d \bx} - \frac{1}{8 \pi} \int_{\Omega} |\nabla \phi(\bx)|^2 \, \mathrm{d\bx}  \nonumber \\
& + & \int_{\Omega}(u^2(\bx)+ b(\bx,\bR)) \phi(\bx) \, \mathrm{d\bx} - \frac{1}{2}\sum_{J} \int_{\Omega} b_J(\bx,\bR_J) V_J(\bx,\bR_J) \, \mathrm{d\bx} + \mathcal{E}_c^*(\bR)  \,. 
\end{eqnarray}


\subsection{OF-DFT ground-state} \label{Subsec:GroundState}
In the framework described above, the variational problem for determining the ground-state in OF-DFT can be written as 
\begin{equation} \label{Eqn:GroundStateSplit}
\mathcal{E}_0 = \inf_{\bR \in \R^{3M_a}} \mathcal{E}^*(\bR) \,,
\end{equation}
where 
\begin{equation} \label{Eqn:GroundStateElectronic}
\mathcal{E}^*(\bR) = \inf_{u \in \mathcal{X}} \mathcal{E}(u,\bR) = \inf_{u \in \mathcal{X}} \bigg\{ \sup_{\phi \in Y} \mathcal{F}(u,\bR,\phi) + \mu T_{LR}(u) \bigg\} \,.
\end{equation}
Through this decomposition, the ground-state can be ascertained by solving the electronic structure problem in Eqn. \ref{Eqn:GroundStateElectronic} for every configuration of the nuclei encountered during the geometry optimization described by Eqn. \ref{Eqn:GroundStateSplit}. Below, we discuss the solution strategy for both of these simulation components.  

\subsubsection{Electronic structure problem}\label{Subsubsec:ElectronicStructure}
Consider the variational problem in Eqn. \ref{Eqn:GroundStateElectronic} for determining the electronic ground-state. On taking the first variation, we arrive at the Euler-Lagrange equation
\begin{equation} \label{Eqn:EulerLagrange}
\mathcal{H} u(\bx) = \eta u(\bx) \,, \quad \mathcal{H} = -\frac{\lambda}{2} \nabla^2  + \bigg( V_{TF}(\bx) + \mu V_{LR}(\bx) + V_{xc}(\bx) + \phi(\bx) \bigg) \,,
\end{equation}
where $V_{LR}(\bx)$ is as given by Eqn. \ref{Eqn:VLinearResponse:local}, and $\eta$ is the Lagrange multiplier used to enforce the constraint $\mathcal{C}(u)=0$. Further, $\phi(\bx)$ is the solution of the Poisson equation 
\begin{equation}\label{Eqn:Poisson}
\frac{-1}{4\pi} \nabla^2 \phi(\bx) = u^{2}(\bx) + b(\bx,\bR) 
\end{equation}
under periodic boundary conditions and
\begin{eqnarray}
V_{TF}(\bx) & = & \frac{\delta T_{TF}(u)}{\delta u^2} = \frac{5}{3} C_F u^{4/3}(\bx) \,, \\
V_{xc}(\bx) & = & \frac{\delta E_{xc}(u)}{\delta u^2} = V_{x}(\bx) + V_{c}(\bx) \,.
\end{eqnarray}
The exchange-correlation potential $V_{xc}(\bx)$ can be decomposed as  
\begin{eqnarray}
V_{x}(\bx) & = & -\left(\frac{3}{\pi}\right)^{1/3}u^{2/3}(\bx) \,, \\
V_{c}(\bx) & = & 
\begin{cases}
\frac{\gamma_1 + \frac{7}{6}\gamma_1 \beta_1 \sqrt{r_s(\bx)} + \frac{4}{3} \gamma_1 \beta_2 r_s(\bx)}{(1+\beta_1\sqrt{r_s(\bx)}+\beta_2{r_s(\bx)})^2}\,, \,\,\,\,\,\,\,\,\,\,\,\
r_s(\bx) \geq 1\\
\left( A_1 + \frac{2}{3} C_1 r_s(\bx) \right) \log r_s(\bx) + \left(B_1 - \frac{1}{3}A_1 \right) + \frac{1}{3} (2D_1-C_1)r_s(\bx)\,, \,\,\,\,\,\, r_s(\bx)<1
\end{cases} 
\end{eqnarray}
with $V_{x}(\bx)$ and $V_{c}(\bx)$ being the exchange and correlation potentials, respectively. Even though the notation does not make it explicit, the dependence of $\mathcal{H}$ on $u$ makes Eqn. \ref{Eqn:EulerLagrange} a non-linear problem. It is worth noting that since $\int_{\Omega} (u^2(\bx)+b(\bx,\bR)) \, \mathrm{d\bx}=0$, the Poisson problem defined by Eqn. \ref{Eqn:Poisson} with periodic boundary conditions is well-posed.

The electronic ground-state can be determined by solving the non-linear eigenvalue problem in Eqn. \ref{Eqn:EulerLagrange} for the eigenfunction corresponding to the lowest eigenvalue. Irrespective of the solution technique and choice of kinetic energy functional, $\phi(\bx)$ needs to be recalculated for every update in $u(\bx)$. The same is true for $V_{LR}(\bx)$ when linear-response kinetic energy functionals are employed. Therefore, the solution of Eqn. \ref{Eqn:EulerLagrange} requires the repeated solution of the Poisson equation in Eqn. \ref{Eqn:Poisson} and the complex-valued non-Hermitian Helmholtz equations in Eqns. \ref{Eqn:Helmholtz:beta} and \ref{Eqn:Helmholtz:alpha}. In view of this, the Self-Consistent Field method (SCF) \cite{fang2009two}---commonly utilized in DFT calculations---is an attractive choice because relatively few iterations are typically required for convergence. However, we have found such an approach to be unstable for both the TFW and WGC kinetic energy functionals, especially as the system size is increased. Since the number of Helmholtz equations that need to be solved can be significantly large in practice (e.g. fifty-two in this work), they are expected to completely dominate the execution time. In order to mitigate this, we develop a fixed-point method for determining the electronic ground-state when linear response kinetic energy functionals are employed \cite{priv:Vikram:2014}. This is similar in spirit to the SCF method, and is found to converge rapidly, as demonstrated by the examples in Section \ref{Section:Examples}.

\begin{figure}[H]
\centering
\includegraphics[keepaspectratio=true,width=1\textwidth]{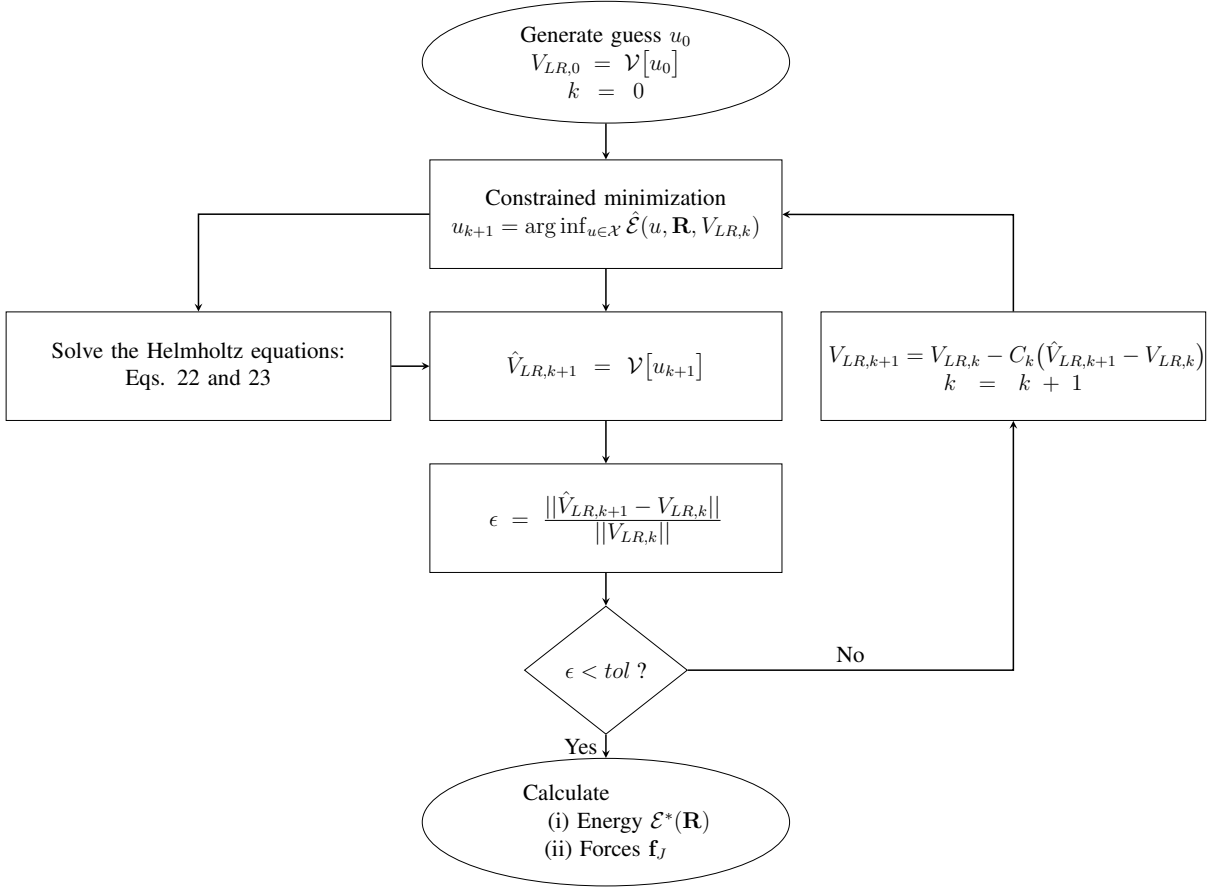}
\caption{Fixed-point iteration for determining the electronic ground state in OF-DFT when linear-response kinetic energy functionals are employed. The functional $\hat{\mathcal{E}}(u,\bR,V_{LR,k}) = \{ \sup_{\phi \in Y} \mathcal{F}(u,\bR,\phi) + \int_{\Omega} V_{LR,k}(\bx) u^2(\bx) \, \mathrm{d\bx} \}$.}
\label{fig:Flowchart}
\end{figure}

We rewrite the nonlinear eigenvalue problem in Eqn. \ref{Eqn:EulerLagrange} as a fixed-point problem with respect to $V_{LR}(\bx)$: 
\begin{equation}
V_{LR} = \mathcal{V} \big[\mathcal{U}(V_{LR}) \big] \,, \label{Eqn:FixedPoint:Map1} 
\end{equation} 
where the mappings
\begin{equation} \label{Eqn:MVF}
\mathcal{U}(V_{LR}) = \arg \inf_{u \in \mathcal{X}} \bigg\{ \sup_{\phi \in Y} \mathcal{F}(u,\bR,\phi) + \mu \int_{\Omega} V_{LR}(\bx) u^2(\bx) \, \mathrm{d\bx} \bigg\} \,,
\end{equation}
and 
\begin{eqnarray}
\mathcal{V}\big [u \big ] & = & C_F \sum_{m=0}^{L} \sum_{n=0}^{L} \sum_{p=0}^m \sum_{q=0}^n  \sum_{r=1}^R  C_{mnpq} \bigg[ (m-p+\alpha)  u^{2(m-p+\alpha-1)}(\bx) V_{mnq\beta r}(\bx) \, \nonumber \\ & + & (n-q+\beta) u^{2(n-q+\beta-1)}(\bx) V_{mnp\alpha r}(\bx) \Bigg] \,.
\end{eqnarray}
Above, $V_{mnq\beta r}(\bx)$ and $V_{mnp\alpha r}(\bx)$ are solutions to the Helmholtz equations given in Eqns. \ref{Eqn:Helmholtz:beta} and \ref{Eqn:Helmholtz:alpha}, respectively. The mapping $\mathcal{U}(V_{LR})$ corresponds to the solution of the nonlinear eigenvalue problem in Eqn. \ref{Eqn:EulerLagrange} for a fixed kernel potential $V_{LR}(\bx)$. The mapping $\mathcal{V}\big [u \big ]$ corresponds to the calculation of $V_{LR}(\bx)$ for some given $u(\bx)$. Therefore, the fixed-point of the composite mapping $\mathcal{V} \big[\mathcal{U}(V_{LR}) \big]$ coincides with the solution of the Euler-Lagrange equation (Eqn. \ref{Eqn:EulerLagrange}) for the electronic ground-state. In order to solve this fixed-point problem, we treat it as a non-linear equation and adopt an iteration of the form \cite{fang2009two,lin2013elliptic}
\begin{equation}
V_{LR,k+1} = V_{LR,k} - C_k \left( \mathcal{V}\big[ \mathcal{U}(V_{LR,k})\big] - V_{LR,k} \right)\,,
\end{equation}
where the index $k$ represents the iteration number and $C_k$ is appropriately chosen to ensure/accelerate convergence. Once the fixed-point $V_{LR}^*(\bx)$ has been determined, $u^*(\bx)$ can be calculated by solving Eqn. \ref{Eqn:MVF} for $V_{LR}(\bx) = V_{LR}^*(\bx)$. In Fig. \ref{fig:Flowchart}, we present a flowchart that outlines the aforedescribed fixed-point approach. It is worth noting that for the choice of TFW kinetic energy functional ($\mu=0$), the solution of Eqn. \ref{Eqn:MVF} coincides with the electronic ground-state. 

After determining the electronic ground-state, the corresponding energy can be evaluated using the expression
\begin{eqnarray} \label{Eqn:GroundStateEnergy}
\mathcal{E}^*(\bR) &=& C_F \int_{\Omega} u^{*10/3} (\bx) \, \mathrm{d\bx} + \frac{\lambda}{2} \int_{\Omega} |\nabla u^{*}(\bx)|^2 \, \mathrm{d\bx} \nonumber \\ 
&+& \frac{\mu}{2}C_F \sum_{m=0}^{L} \sum_{n=0}^{L} \sum_{p=0}^m \sum_{q=0}^n  \sum_{r=1}^R C_{mnpq} \int_{\Omega} \bigg[ u^{*2(m-p+\alpha)}(\bx) V^*_{mnq\beta r}(\bx)   +  u^{*2(n-q+\beta)}(\bx) V^*_{mnp\alpha r}(\bx) \bigg] \, \mathrm{d\bx} \nonumber \\
&+& \int_{\Omega} \varepsilon_{xc} (u^*(\bx)) u^{*2}(\bx) \, \mathrm{d \bx} + \frac{1}{2} \int_{\Omega}(u^{*2}(\bx)+ b(\bx,\bR)) \phi^{*}(\bx) \, \mathrm{d\bx}  \nonumber \\ 
&-& \frac{1}{2}\sum_{J} \int_{\Omega} b_J(\bx,\bR_J) V_J(\bx,\bR_J) \, \mathrm{d\bx} + \mathcal{E}_c^*(\bR)   \,,
\end{eqnarray}
where $V^*_{mnq\beta r}(\bx)$, $V^*_{mnp\alpha r}(\bx)$ and $\phi^*(\bx)$ are solutions of Eqns. \ref{Eqn:Helmholtz:beta}, \ref{Eqn:Helmholtz:alpha} and \ref{Eqn:Poisson}, respectively, for $u(\bx) = u^*(\bx)$.

\subsubsection{Geometry optimization: forces on nuclei}\label{Subsec:GeometryOptimization}
Consider the minimization problem in Eqn. \ref{Eqn:GroundStateSplit} for determining the equilibrium configuration of the atoms. During this geometry optimization, the forces on the nuclei can be calculated using the relation 
\begin{eqnarray}\label{Eqn:Force:Nuclei}
\mathbf{f}_J & = & -\frac{\partial \mathcal{E}^*(\bR)}{\partial \bR_J}  \nonumber \\
 & = & - \sum_{J'} \int_{\Omega} \frac{\partial b_{J'}(\bx,\bR_{J'})}{\partial \bR_{J'}}\left(\phi^*(\bx)-V_{J'}(\bx,\bR_{J'})\right) \, \rm{d\bx} + \mathbf{f}_J^c \,, \label{1Eqn:Force:Nuclei} \\
 & = & \sum_{J'} \int_{\Omega} \nabla b_{J'}(\bx,\bR_{J'}) \left(\phi^*(\bx)-V_{J'}(\bx,\bR_{J'})\right) \, \rm{d\bx} + \mathbf{f}_J^c \,, \nonumber
\end{eqnarray}
where ${\bf f}_J$ denotes the force on the $J^{th}$ nucleus and the summation over $J'$ signifies the $J^{th}$ atom and its periodic images. Additionally, $\phi^*(\bx)$ is the solution of the Poisson equation in Eqn. \ref{Eqn:Poisson} for $u(\bx)=u^*(\bx)$ and ${\bf f}_J^c = -\frac{\partial \mathcal{E}_c^*(\bR)}{\partial \bR_J}$ corrects for the error in forces due to overlapping charge density of nuclei. The expression for this correction has been derived in Appendix \ref{Appendix:Correct:RepulsiveEnergy}. The second equality in Eqn. \ref{Eqn:Force:Nuclei} is obtained by using the fact that the energy is stationary with respect to $u(\bx)$ and $\phi(\bx)$ at the electronic ground-state, and the last equality is obtained by using the spherical symmetry of $b_{J'}(\bx,\bR_{J'})$ (i.e., $b_{J'}(\bx,\bR_{J'}) \equiv b_{J'}(|\bx-\bR_{J'} |)$). Since $\nabla b_{J'}(\bx,\bR_{J'})$ has compact support in a ball of radius $r_c$ centered at $\bR_{J'}$, only a finite number of periodic images of the $J^{th}$ atom have an overlap with $\Omega$. Therefore, evaluation of the atomic forces is amenable to a linear-scaling real-space implementation. 


\section{Numerical Implementation} \label{Section:NumericalImplementation}
In this section, we describe a higher-order finite-difference implementation of the formulation presented in the previous section. We restrict our computation to a cuboidal domain $\Omega$ of sides $L_1$, $L_2$ and $L_3$. We generate a uniform finite-difference grid with spacing $h$ such that $L_1=n_1 h$, $L_2=n_2h$ and $L_3=n_3h$, where $n_1$, $n_2$ and $n_3$ are natural numbers. We index the grid points by $(i,j,k)$, where $i=1,2,\ldots, n_1$, $j=1,2,\ldots, n_2$ and $k=1,2,\ldots, n_3$. We approximate the Laplacian of any function $f(\bx)$ at the grid point $(i,j,k)$ using higher-order finite-differences \cite{LeVeque2007}
\begin{eqnarray}\label{Eqn:FD:Laplacian}
\nabla^2 f \big|^{(i,j,k)} \approx \sum_{p=0}^{N} w_p \bigg(f^{(i+p,j,k)} + f^{(i-p,j,k)} + f^{(i,j+p,k)} + f^{(i,j-p,k)} + f^{(i,j,k+p)} + f^{(i,j,k-p)} \bigg) \,,
\end{eqnarray}
where $f^{(i,j,k)}$ represents the value of the function $f(\bx)$ at the grid point $(i,j,k)$.  The weights $w_p$ are given by \cite{mazziotti1999spectral,jordan2003spectral,Suryanarayana2014524}
\begin{eqnarray}
w_0 & = & - \frac{1}{h^2} \sum_{q=1}^N \frac{1}{q^2} \,, \nonumber \\
w_p & = & \frac{2 (-1)^{p+1}}{h^2 p^2} \frac{(N!)^2}{(N-p)! (N+p)!} \,, \,\, p=1, 2, \ldots, N.
\end{eqnarray} 
Similarly, we approximate the gradient at the grid point $(i,j,k)$ using higher-order finite-differences
\begin{eqnarray}\label{Eqn:gradient:approximate}
\nabla f \big|^{(i,j,k)} \approx \sum_{p=1}^{N} \tilde{w}_p \bigg( ( f^{(i+p,j,k)} - f^{(i-p,j,k)}) \hat{\mathbf{e}}_1 + ( f^{(i,j+p,k)} - f^{(i,j-p,k)}) \hat{\mathbf{e}}_2  + ( f^{(i,j,k+p)} - f^{(i,j,k-p)}) \hat{\mathbf{e}}_3 \bigg) \,,
\end{eqnarray}
where $\hat{\mathbf{e}}_1$, $\hat{\mathbf{e}}_2$ and $\hat{\mathbf{e}}_3$ represent unit vectors along the edges of the cuboidal domain $\Omega$. The weights $\tilde{w}_p$ are given by \cite{mazziotti1999spectral,jordan2003spectral,Suryanarayana2014524}
\begin{equation}
\tilde{w}_p = \frac{(-1)^{p+1}}{h p} \frac{(N!)^2}{(N-p)! (N+p)!} \,, \,\, p=1, 2, \ldots, N.
\end{equation}
These finite-difference expressions for the Laplacian and gradient represent $2 N$ order accurate approximations, i.e. error is $\mathcal{O}(h^{2N})$. While performing spatial integrations, we assume that the function $f(\bx)$ is constant in a cube of side $h$ around each grid point, i.e.
\begin{equation} \label{Eqn:IntApprox}
\int_{\Omega} f(\bx) \, \mathrm{d\bx} \approx h^3  \sum_{i=1}^{n_1} \sum_{j=1}^{n_2} \sum_{k=1}^{n_3}f^{(i,j,k)} .
\end{equation} 
We enforce periodic boundary conditions on $\Omega$ by employing the following strategy. In the finite-difference representations of the Laplacian and gradient presented in Eqns. \ref{Eqn:FD:Laplacian} and \ref{Eqn:gradient:approximate} respectively, we map any index that does not correspond to a node in the finite-difference grid to its periodic image within $\Omega$. 

We start with precomputed radially-symmetric and compactly-supported isolated-atom electron densities for each type of atom. We superimpose these isolated-atom electron densities for the initial configuration of the nuclei, and scale the resulting electron density such that the constraint on the total number of electrons is satisfied. We take the pointwise square-root of the electron density so obtained as the starting guess $u_0^{(i,j,k)}$. During the aforedescribed calculation, we only visit atoms whose isolated-atom electron densities have non-zero overlap with $\Omega$. Similarly, for every new configuration of atoms encountered during the geometry optimization, we calculate the charge density of the nuclei using the relations 
\begin{equation}
b^{(i,j,k)} = \sum_J b_{J}^{(i,j,k)} \,, \quad b_{J}^{(i,j,k)} = -\frac{1}{4\pi} \nabla^2  V_J \big|^{(i,j,k)} \,,
\end{equation}
where the summation reduces to all atoms whose charge density has non-zero overlap with $\Omega$. The localized nature of the above operations ensures that the evaluation of $u_0^{(i,j,k)}$ and $b^{(i,j,k)}$ scales linearly with the number of atoms. 
  
We solve the variational problem in Eqn. \ref{Eqn:MVF} using a conjugate gradient method that was originally developed for DFT \cite{teter1989solution,payne1992iterative} and later adopted in simplified form for OF-DFT \cite{Teter1992,jiang2004conjugate,Ho2008}. Specifically, we utilize the Polak-Ribiere update \cite{Shewchuk1994} with Brent's method \cite{press2007numerical} for the line-search. We refer the reader to Appendix \ref{Appendix:NLCGTeter} for further details on the implemented algorithm. For every update in the square-root electron density, we solve the Poisson equation in Eqn. \ref{Eqn:Poisson} under periodic boundary conditions using the Generalized minimal residual (GMRES) \cite{saad1986gmres} method with the block-Jacobi preconditioner \cite{golub2012matrix}. Since the solution so obtained is accurate to within an indeterminate constant, we enforce the condition $\int_{\Omega} \phi(\bx) \, \mathrm{d\bx}=0$ for definiteness. In every subsequent Poisson equation encountered, we use the previous solution as starting guess. For the complex-valued Helmholtz equations in Eqns. \ref{Eqn:Helmholtz:beta} and \ref{Eqn:Helmholtz:alpha}, we first separate out each equation into its real and imaginary parts, and then solve the resulting coupled equations simultaneously under periodic boundary conditions using GMRES with block-Jacobi preconditioners. In every iteration of the fixed-point method, we use the solution of the Helmholtz equations from the previous iteration as the starting guess. We accelerate the convergence of the fixed-point iteration by utilizing Anderson mixing \cite{anderson1965iterative}, details of which can be found in Appendix \ref{Appendix:Anderson}.  

Once the electronic ground-state square-root electron density has been determined, the energy and forces are evaluated using Eqns. \ref{Eqn:GroundStateEnergy} and \ref{Eqn:Force:Nuclei} respectively. While doing so, we restrict the summation over the periodic images to atoms whose charge densities have non-zero overlap with $\Omega$. We solve for the equilibrium configuration of the atoms by using the conjugate gradient method with the Polak-Ribiere update and secant line search \cite{Shewchuk1994}. We have developed a parallel implementation of the proposed approach using the Portable, Extensible Toolkit for scientific computations (PETSc) \cite{Petsc1,Petsc2} suite of data structures and routines. Within PETSc, we have utilized distributed arrays with the star-type stencil option. The communication between the processors is handled via the Message Passing Interface (MPI) \cite{gropp1999using}. 


\section{Examples and Results} \label{Section:Examples}
In this section, we validate the proposed formulation and higher-order finite-difference implementation of periodic OF-DFT through selected examples. Henceforth, we shall refer to this framework as RS-FD, which is an acronym for Real-Space Finite-Differences. In all the simulations, we employ the Goodwin-Needs-Heine pseudopotential \cite{goodwin1990pseudopotential}. In addition, we choose $\lambda=\frac{1}{5}$ for the TFW functional, and $\lambda=1$, $L=2$, $R=4$, $\alpha=\frac{5}{6}+ \frac{\sqrt{5}}{6}$ and $\beta=\frac{5}{6} - \frac{\sqrt{5}}{6}$ for the WGC functional. Wherever applicable, we compare our results with the plane-wave code PROFESS \cite{Ho2008,Hung2010}. Within PROFESS, we utilize a plane-wave energy cutoff of $E_{cut}=1200$ eV, which results in energies and forces that are converged to within $1\times 10^{-6}$ eV/atom and $6 \times 10^{-4}$ eV/Bohr respectively. Unless specified otherwise, we use sixth-order accurate finite-differences and a mesh size of $h=0.5$ Bohr within RS-FD. We choose a cutoff radius of $10$ Bohr for the isolated-atom electron densities as well as the charge densities of the nuclei, whereby the enclosed charge for each nucleus is accurate to within $5 \times 10^{-9}$. We utilize tolerances of $1 \times 10^{-7}$ and $1 \times 10^{-12}$ on the normalized residual as the stopping criterion for the conjugate gradient and GMRES methods, respectively. We employ a history of $m=3$ in Anderson mixing and a tolerance of $1 \times 10^{-7}$ on the normalized residual for convergence of the fixed-point method. These parameters and tolerances result in RS-FD energies and forces that are converged to within $0.007$ eV/atom and $0.007$ eV/Bohr, respectively. It is worth noting that the aforementioned RS-FD tolerances are highly conservative, i.e. chemical accuracies are achieved even when they are significantly relaxed, as discussed in Section \ref{Subsec:ScalingPerformance}.  We perform all simulations on computer cluster wherein each node has the following configuration: Altus 1804i Server - 4P Interlagos Node, Quad AMD Opteron 6276, 16C, 2.3 GHz, 128GB, DDR3-1333 ECC, 80GB SSD, MLC, 2.5" HCA, Mellanox ConnectX 2, 1-port QSFP, QDR, memfree, CentOS, Version 5, and connected through InfiniBand cable.


\subsection{Convergence of energy with spatial discretization} \label{Subsec:ConvergenceEnergyMesh}
\begin{figure}[H]
\centering
\subfloat[TFW]{\label{fig:EnergyConvTFW}\includegraphics[keepaspectratio=true,width=0.45\textwidth]{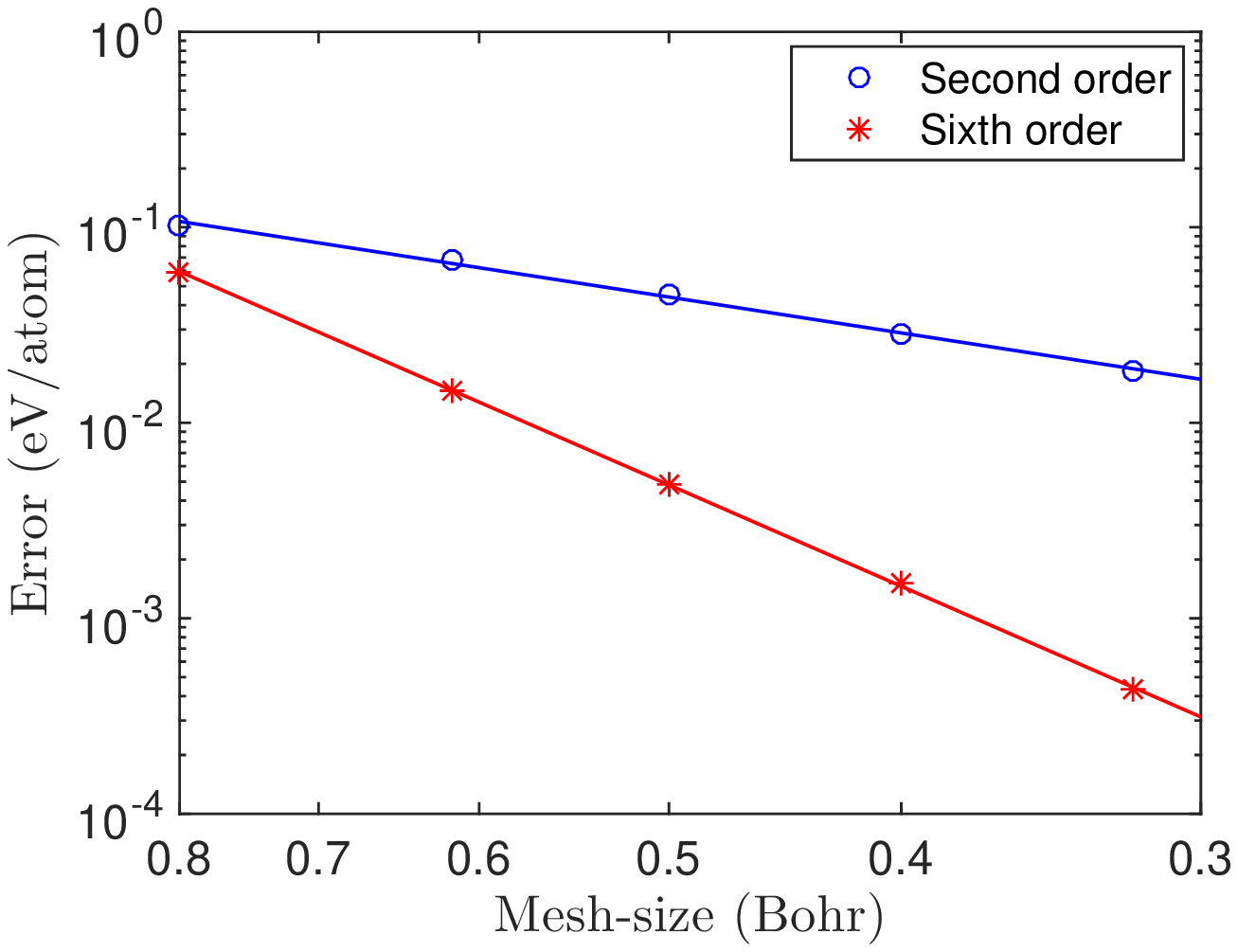}}
\subfloat[WGC]{\label{fig:EnergyConvWGC}\includegraphics[keepaspectratio=true,width=0.45\textwidth]{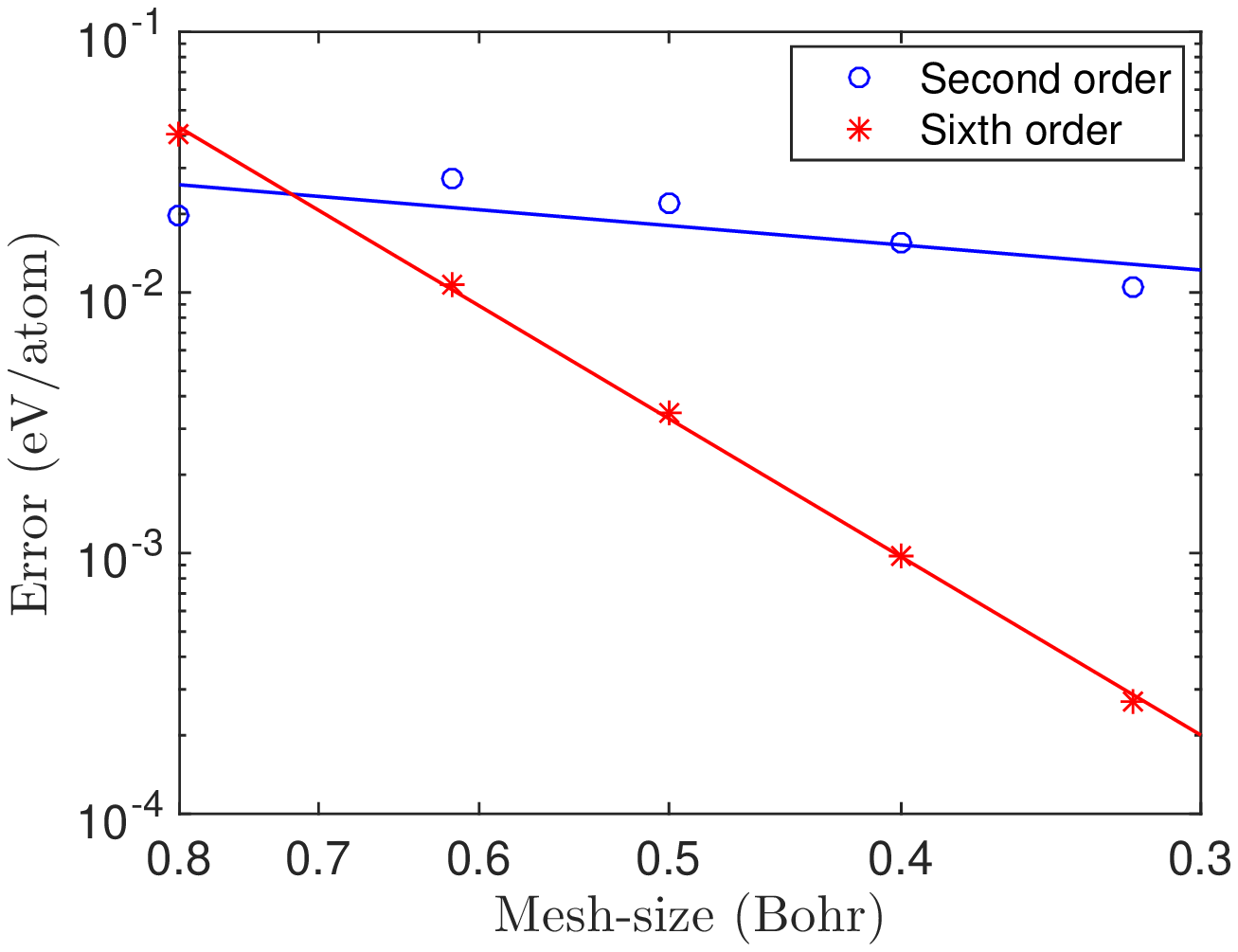}} \\
\caption{Convergence of the energy with respect to the finite-difference mesh-size ($h$). The reference energy is computed using sixth-order finite-differences with $h=0.16$ Bohr. } 
\label{Fig:ConvergenceEnergy}
\end{figure}

We start by verifying convergence of the energy computed by RS-FD with respect to the mesh-size ($h$). As the representative example, we choose a $4$-atom Face-Centered Cubic (FCC) unit cell of Aluminum with lattice constant of $a=8.0$ Bohr, and displace the atom at the corner of the unit cell---the origin of the coordinate system---to [$0.80$ $0.56$ $0.42$] Bohr. We evaluate the energy of this system as a function of $h$ for second and sixth-order accurate finite-difference approximations. In Fig. \ref{Fig:ConvergenceEnergy}, we plot the resulting convergence in energy for the TFW and WGC kinetic energy functionals, with the reference value computed using sixth-order finite-differences and $h=0.16$ Bohr. We observe that sixth-order finite-differences demonstrates significantly higher rates of convergence compared to second-order finite-differences. Specifically, the sixth-order scheme obtains convergence rates of $5.35$ and $5.47$ for the TFW and WGC functionals, respectively, whereas the second-order discretization obtains rates of $1.90$ and $0.77$, respectively. Interestingly, the computed convergence rates are not equal to the order of the finite-difference approximation. Possible reasons for this include the nonlinearity of the problem, need for finer meshes to obtain the asymptotic convergence rates, the use of trapezoidal rule for integration, and the ``egg-box" effect. Overall, these results indicate that second-order finite-differences are prohibitively expensive for obtaining the chemical accuracies desired in OF-DFT calculations, thereby motivating higher-order approximations.  

\subsection{Convergence of atomic forces with spatial discretization} \label{Subsec:ConvergenceForcesMesh}
\begin{figure}[H]
\centering
\subfloat[TFW]{\label{fig:ForceConvTFW}\includegraphics[keepaspectratio=true,width=0.45\textwidth]{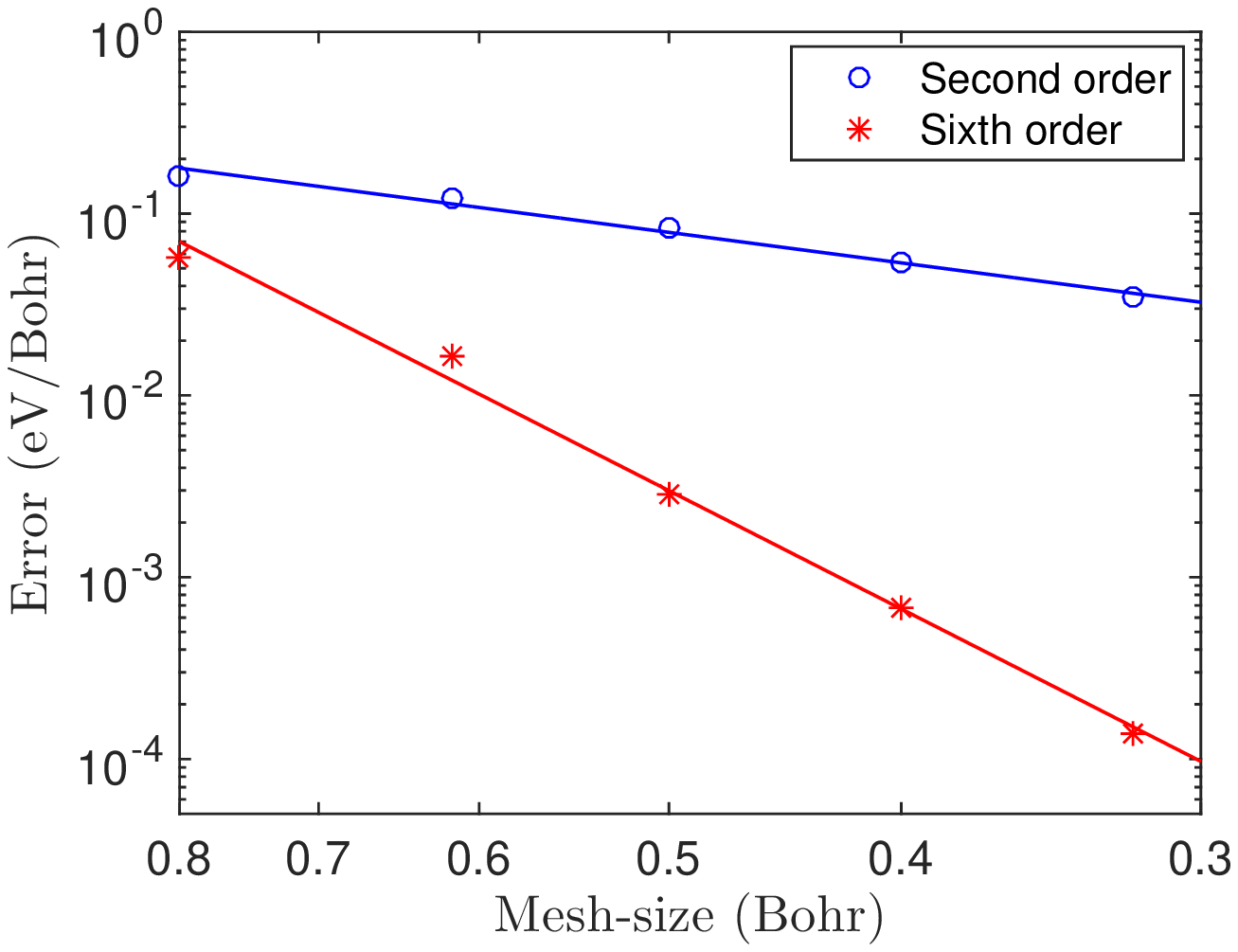}}
\subfloat[WGC]{\label{fig:ForceConvWGC}\includegraphics[keepaspectratio=true,width=0.45\textwidth]{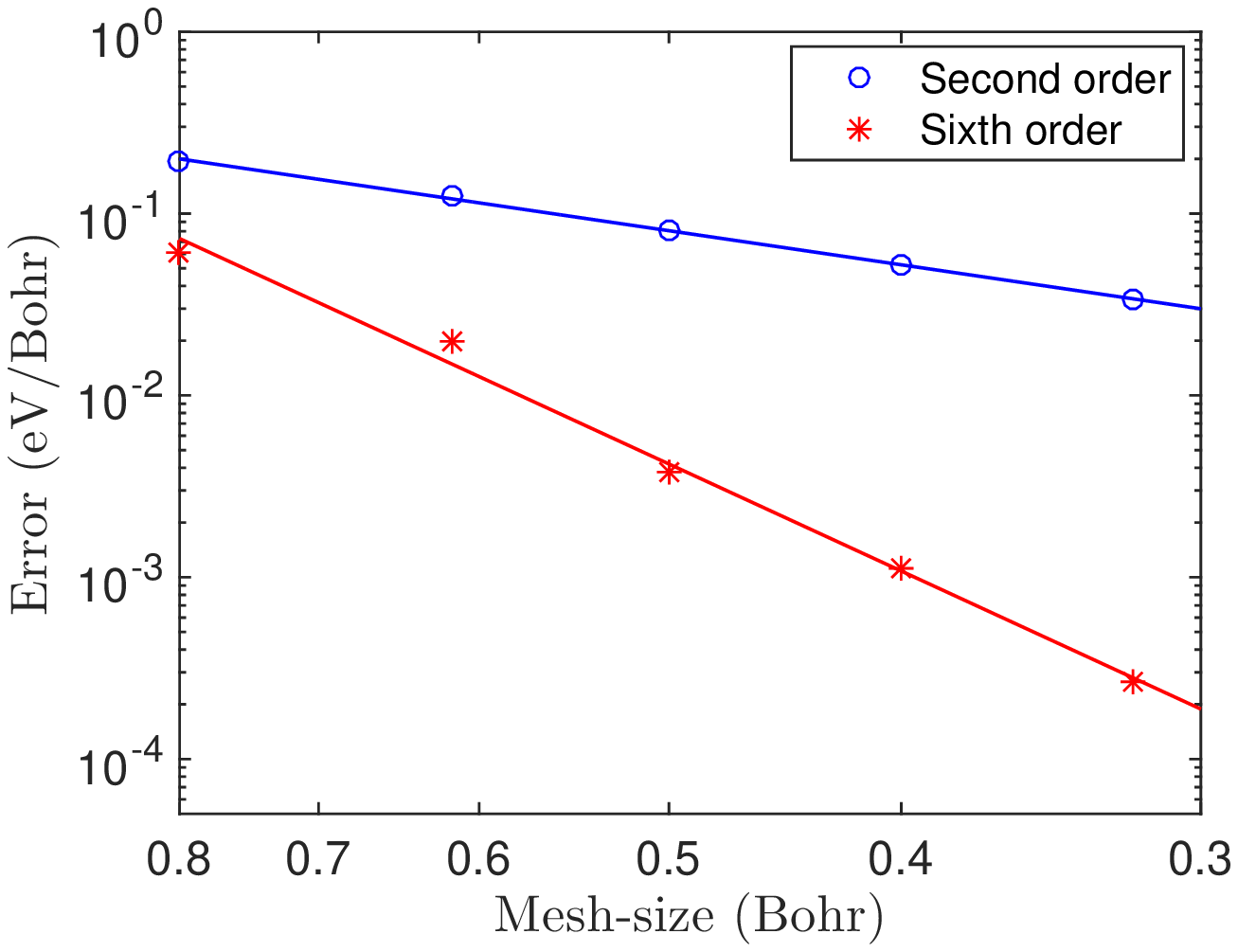}} \\
\caption{Convergence of the atomic force with respect to the finite-difference mesh-size ($h$). The reference force is computed using sixth-order finite-differences with mesh-size of $h=0.16$ Bohr.} 
\label{Fig:ConvergenceForce}
\end{figure}

Next, we verify the convergence of the atomic forces with respect to the mesh-size ($h$). We choose the same example as that used for studying convergence of the energy in Section \ref{Subsec:ConvergenceEnergyMesh}. We calculate the force on the displaced atom for the TFW and WGC kinetic energy functionals, and plot the resulting error versus $h$ in Fig. \ref{Fig:ConvergenceForce}. The error is defined to be the maximum difference in the force from that obtained using sixth-order finite-differences with mesh-size of $h=0.16$ Bohr. We again observe that sixth-order finite-differences demonstrates significantly larger convergence rates compared to second-order finite-differences. Specifically, the convergence rates obtained by the sixth-order scheme for TFW and WGC are $6.71$ and $6.07$, respectively, whereas the rates for the second-order approximation are $1.73$ and $1.93$, respectively. Notably, the convergence rates for the force are larger than those obtained for the energy when using a sixth-order discretization.  The possible reasons for the convergence rates not matching the finite-difference order are the need for finer meshes for obtaining asymptotic rates, the non-variational nature of the finite-difference approximation, and the ``egg-box" effect. 

Overall, we conclude from the results presented in the previous and current subsection that higher-order finite-differences are necessary for performing accurate and efficient electronic structure calculations based on OF-DFT. Indeed, larger convergence rates may be possible as the order of the finite-difference approximation is increased. However, this comes at the price of increased computational cost per iteration due to the reduced locality of the discretized operators and larger inter-processor communication. We have found sixth-order finite-differences to be an efficient choice, which is in agreement with our previous conclusions for the non-periodic TFW setting \cite{Suryanarayana2014524}. In view of this, we will employ sixth-order finite-differences for all the remaining simulations in this work. 


\subsection{Convergence of the fixed-point method} \label{Subsubsec:AndersonMixing}
\begin{figure}[H]
\centering
\subfloat[Perfect crystal]{\label{fig:AndersonPerfect:Param}\includegraphics[keepaspectratio=true,width=0.45\textwidth]{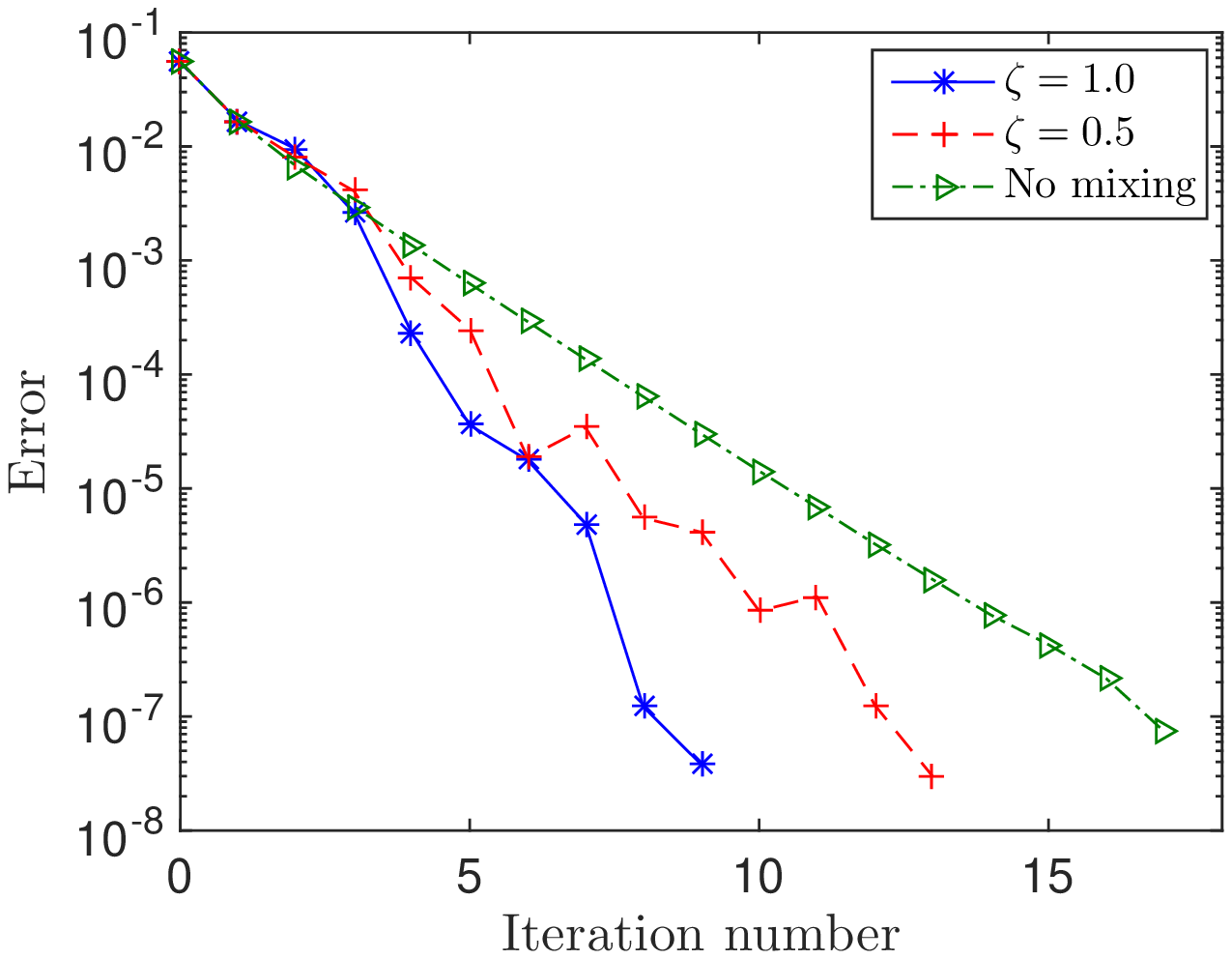}}
\subfloat[Vacancy]{\label{fig:AndersonVacancy:Param}\includegraphics[keepaspectratio=true,width=0.45\textwidth]{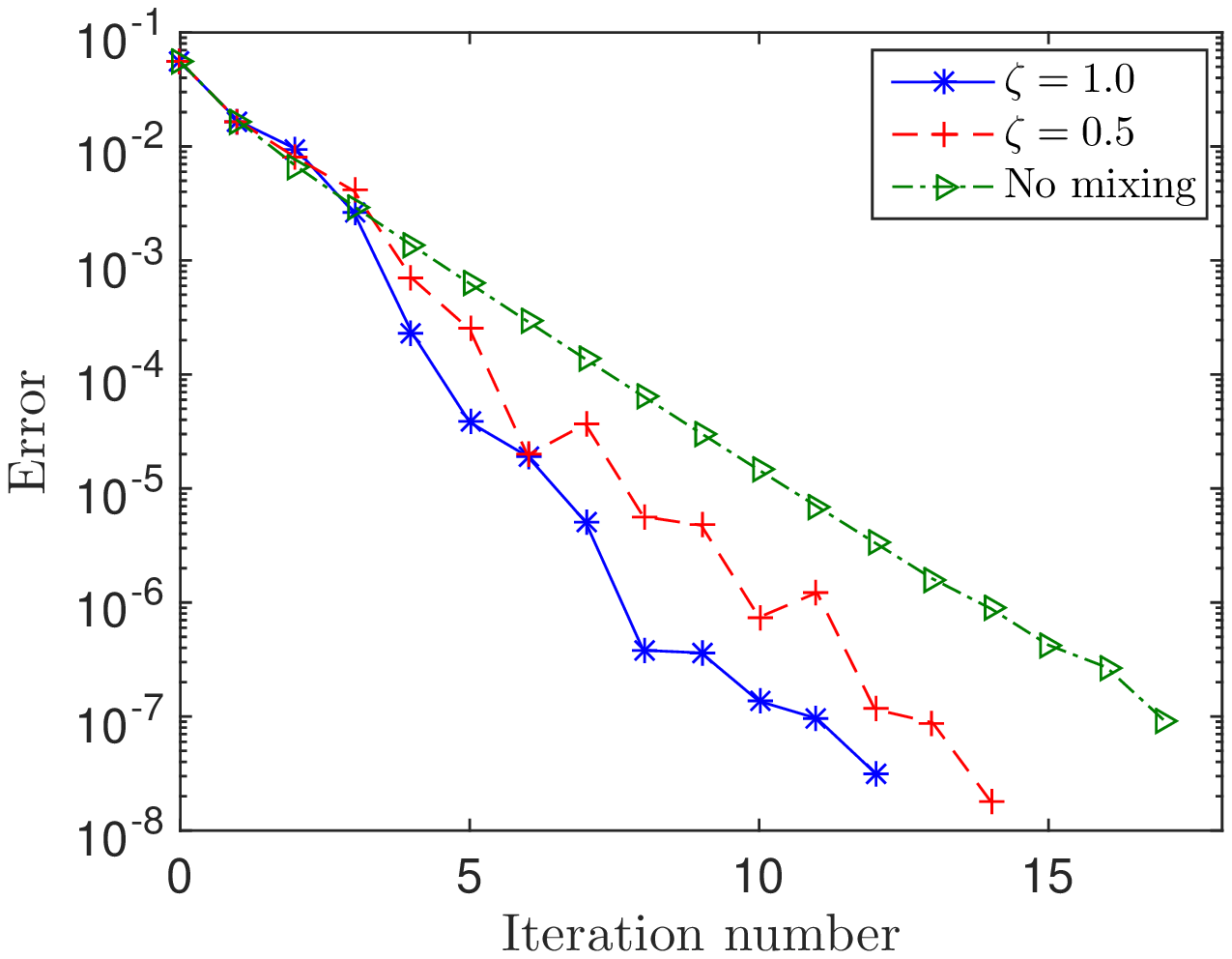}} \\
\caption{Comparison of convergence in the fixed-point iteration with and without Anderson mixing. The mixing history size $m=3$. The error is defined to be the normalized residual $\norm{\hat{V}_{LR,k+1} - V_{LR,k}}/\norm{V_{LR,k}}$, where $k$ denotes the iteration number. The system under consideration is $6 \times 6 \times 6$ FCC Aluminum unit cells with lattice constant of $a=7.50$ Bohr.  }
\label{Fig:AndersonMixing:MixParam}
\end{figure}

We now demonstrate convergence of the fixed-point method for simulations involving the WGC kinetic energy functional. For this study, we choose (i) $864$-atom system consisting of $6 \times 6 \times 6$ FCC unit cells of Aluminum with lattice constant of $a=7.50$ Bohr (ii) $863$-atom system consisting of a vacancy in $6 \times 6 \times 6$ FCC unit cells of Aluminum with lattice constant of $a=7.50$ Bohr. For these two examples, we plot in Fig. \ref{Fig:AndersonMixing:MixParam} the progression of error during the fixed-point iteration. Specifically, we compare the convergence of the basic fixed-point iteration (i.e. no mixing) with that accelerated by Anderson mixing. Within Anderson mixing, we choose mixing history size $m=3$, and mixing parameters $\zeta=1.0$ and $\zeta=0.5$. We observe that Anderson mixing significantly accelerates the convergence of the fixed-point iteration, with the mixing parameter $\zeta=1$ demonstrating the best performance. We have found these results to be representative of other calculations utilizing the WGC kinetic energy functional. In view of this, we will utilize Anderson mixing with mixing parameter $\zeta=1$ for the remaining simulations in this work.
 
\begin{figure}[H]
\centering
\subfloat[Perfect crystal]{\label{fig:AndersonPerfect:Hist}\includegraphics[keepaspectratio=true,width=0.45\textwidth]{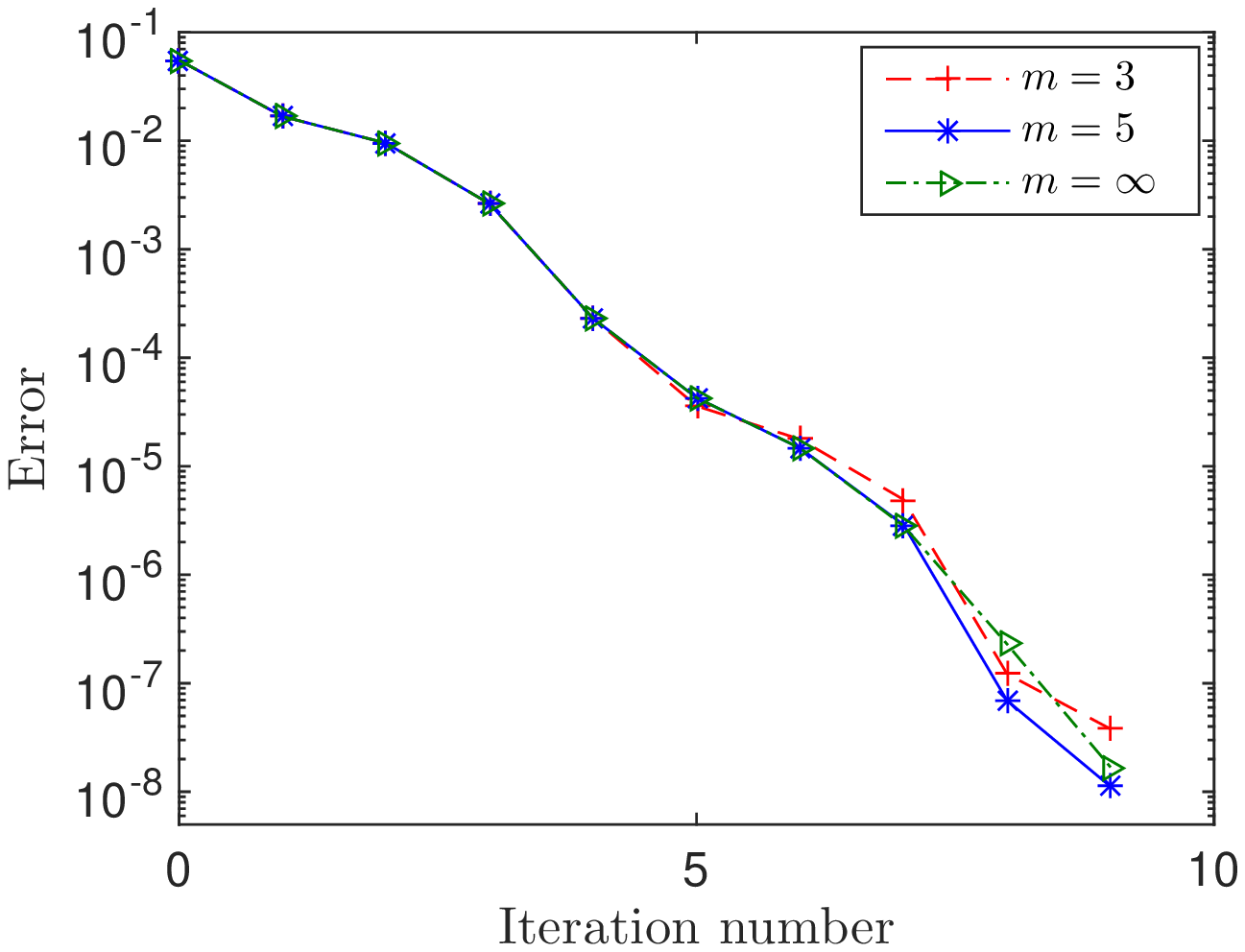}}
\subfloat[Vacancy]{\label{fig:AndersonVacancy:Hist}\includegraphics[keepaspectratio=true,width=0.45\textwidth]{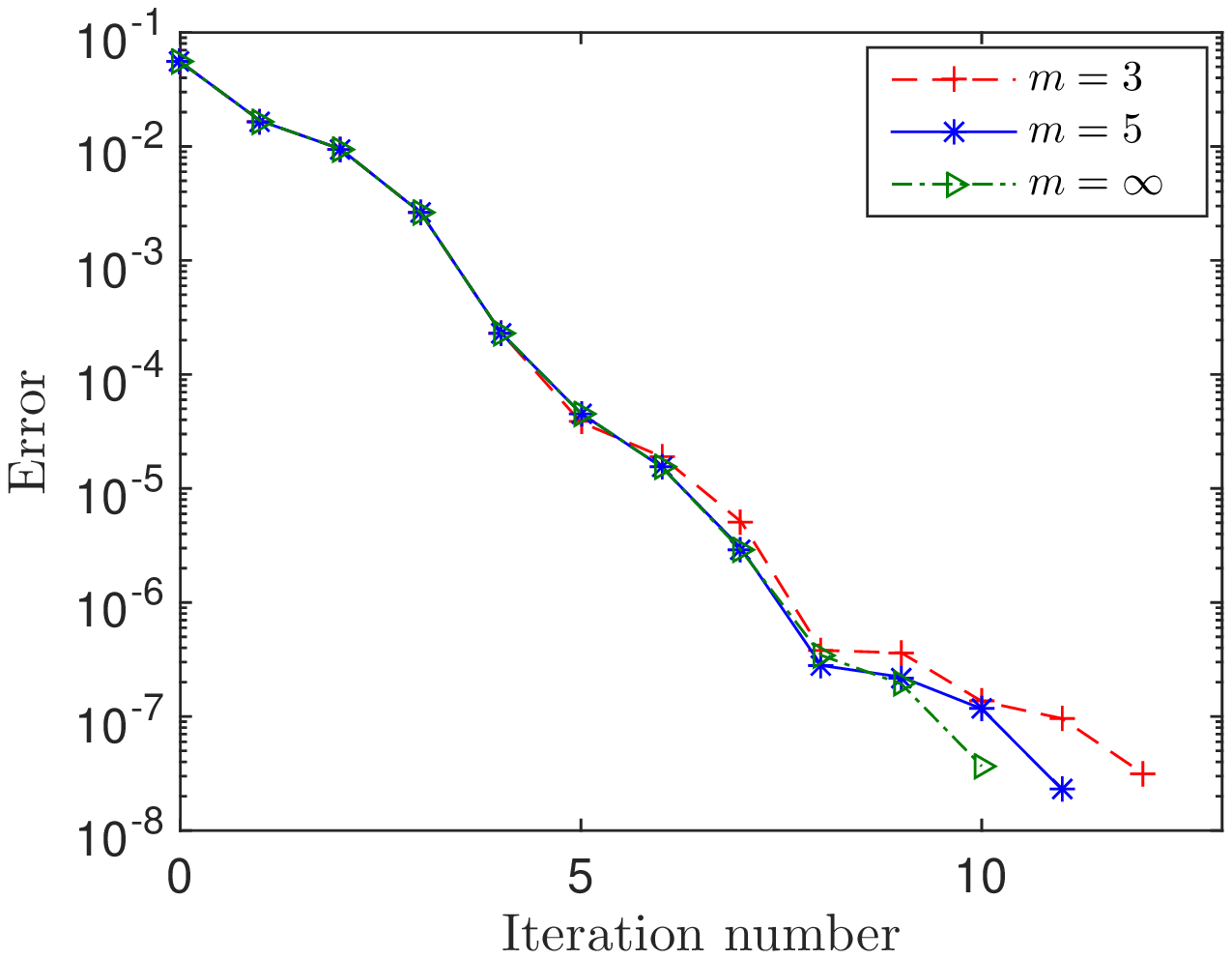}} \\
\caption{Comparison of convergence in the fixed-point iteration for different sizes of mixing history ($m$). The mixing parameter $\zeta=1$. The error is defined to be the normalized residual $\norm{\hat{V}_{LR,k+1} - V_{LR,k}}/\norm{V_{LR,k}}$, where $k$ denotes the iteration number. The system under consideration is $6 \times 6 \times 6$ FCC Aluminum unit cells with lattice constant of $a=7.50$ Bohr. } 
\label{Fig:AndersonMixing:MixHist}
\end{figure}

In Fig. \ref{Fig:AndersonMixing:MixHist}, we compare the convergence of the Anderson accelerated fixed-point iteration for  mixing histories of different sizes. Specifically, we choose $m=3$, $m=5$, and $m=\infty$ for this study. We observe that the size of the mixing history does not have any noticeable impact on the fixed-point iteration. In fact, the plots of the error versus iteration number in Fig. \ref{Fig:AndersonMixing:MixHist} are nearly identical. Overall, we conclude that the fixed-point iteration accelerated with Anderson mixing is extremely robust and efficient. In particular, the error decreases rapidly, and approximately $5$ iterations are sufficient to obtain the desired chemical accuracy in energies and forces. Indeed, the energies are converged to within $1 \times 10^{-6}$ eV/atom and the forces are converged to within $5 \times 10^{-4}$ eV/Bohr for a fixed-point iteration error of $1 \times 10^{-3}$ in Figs. \ref{Fig:AndersonMixing:MixParam} and \ref{Fig:AndersonMixing:MixHist}. 

In this work, we have proposed a fixed-point problem with respect to $V_{LR}(\bx)$ for simulations involving linear-response kinetic energy functionals. However, it is also possible to develop an analogous fixed-point problem with respect to $u(\bx)$. In Table \ref{Table:MixComparison}, we compare the performance of the fixed-point iterations with respect to $u(\bx)$ and $V_{LR}(\bx)$ for the WGC functional. In both cases, we accelerate the iteration using Anderson mixing with $m=3$. It is clear that the relative performance of the two fixed-point iterations is system dependent. However, we have found that the iteration with respect to $V_{LR}(\bx)$ is significantly more robust than the one with $u(\bx)$. Therefore, we employ the fixed-point iteration with respect to $V_{LR}(\bx)$ for determining the electronic ground-state in simulations involving linear-response kinetic energy functionals.

\begin{table}[H]
\centering
\begin{tabular}{ccc}
\hline 
System & Fixed-point problem for $u$ & Fixed-point problem for $V_{LR}$ \\ 
\hline
 $3 \times 3 \times 3$ FCC unit cells perfect crystal & $15$ & $22$ \\
 $6 \times 6 \times 6$ FCC unit cells with a vacancy & $29$ & $18$ \\ 
\hline
\end{tabular}
\caption{Number of steps required to reduce the error to $1 \times 10^{-7}$ in the fixed-point iterations with respect to $u$ and $V_{LR}$.  Anderson mixing with $m=3$ has been employed in both cases.}
\label{Table:MixComparison}
\end{table}


\subsection{Examples} \label{Subsec:Examples}

\subsubsection{Aluminum clusters} \label{Subsec:ExamplesClusters}
First, we study Aluminum clusters consisting of $14$, $172$, $666$, $1688$ and $3430$ atoms that are arranged as $1\times1\times1$, $3\times3\times3$, $5\times5\times5$, $7\times7\times7$ and $9\times9\times9$ FCC unit cells, respectively. The atoms are held fixed, with the lattice constants chosen to minimize the energy \cite{Suryanarayana2014}. The size of the cubical domains are such that the minimum distance of any atom to the boundary is $12$ Bohr. In order to avoid the vacuum resulting divergences encountered when using WGC, we only employ the TFW kinetic energy functional. In Tables \ref{Table:AluminumClustersEnergy} and \ref{Table:AluminumClustersForce}, we compare the energies and forces computed by RS-FD with PROFESS. It is clear that there is very good agreement in the energies and forces. In particular, the maximum difference in the energy is $0.005$ eV/atom and the maximum difference in the forces is $0.00683$ eV/Bohr. Indeed, the agreement between RS-FD and PROFESS improves as the mesh is refined from the currently used value of $h=0.5$ Bohr. For example, at $h=0.25$ Bohr, the difference in the energy and force ($sup$ norm) between RS-FD and PROFESS are $1.1 \times 10^{-4}$ eV/atom and $6.4 \times 10^{-4}$ eV/Bohr, respectively.

\begin{table}[H]
\centering
\begin{tabular}{ccccc}
\hline 
No. of FCC & No. of atoms & $a_e$ (Bohr) &  $\mathcal{E}$ (eV/atom) & $\mathcal{E}$ (eV/atom) \\
unit cells & $(M_a)$ & & RS-FD & PROFESS \\
\hline
$1\times1\times1$ & $14$ & $7.73$ & $-59.246$ & $-59.241$  \\
$3\times3\times3$ & $172$ & $7.89$ & $-59.813$ & $-59.808$  \\
$5\times5\times5$ & $666$ & $7.93$ & $-59.965$ & $-59.960$  \\
$7\times7\times7$ & $1688$ & $7.95$ & $-60.035$ & $-60.030$  \\
$9\times9\times9$ & $3430$ & $7.96$ & $-60.075$ & $-60.071$ \\
\hline
\end{tabular}
\caption{Energy of the clusters consisting of $m\times m\times m$ FCC unit cells of Aluminum, where $m=1$, $3$, $5$, $7$ and $9$.}
\label{Table:AluminumClustersEnergy}
\end{table} 

\begin{table}[H]
\centering
\begin{tabular}{cccc}
\hline 
FCC Aluminum & $l_1$ norm/($3M_a$) & $l_2$ norm/($3M_a$) & $sup$ norm  \\
unit cells & (eV/Bohr) & (eV/Bohr) & (eV/Bohr) \\
\hline
$1\times1\times1$ & $0.00070$ & $0.00013$ & $0.00133$ \\
$3\times3\times3$ & $0.00082$ & $0.00004$ & $0.00211$ \\
$5\times5\times5$ & $0.00160$ & $0.00004$ & $0.00409$ \\
$7\times7\times7$ & $0.00038$ & $0.00001$ & $0.00217$ \\
$9\times9\times9$ & $0.00158$ & $0.00002$ & $0.00683$ \\
\hline
\end{tabular}
\caption{Difference in forces from PROFESS for $m\times m\times m$ FCC Aluminum unit cell clusters, where $m=1$, $3$, $5$, $7$ and $9$.}
\label{Table:AluminumClustersForce}
\end{table} 


\subsubsection{Aluminum crystal} \label{Subsubsec:ExamplesPeriodic}
\begin{figure}[H]
\centering
\subfloat[TFW]{\label{fig:EnergyLatticeTFW}\includegraphics[keepaspectratio=true,width=0.45\textwidth]{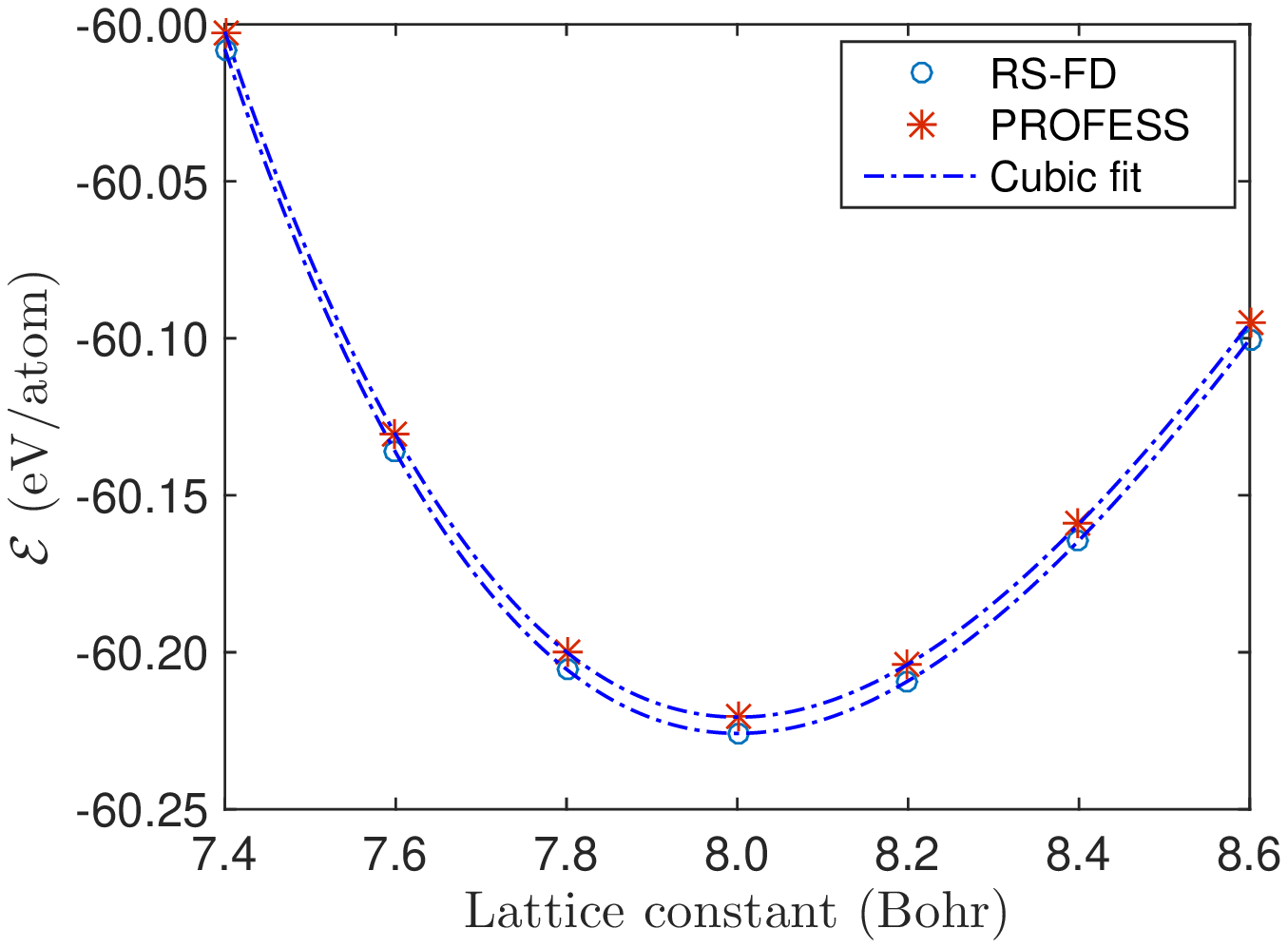}}
\subfloat[WGC]{\label{fig:EnergyLatticeWGC}\includegraphics[keepaspectratio=true,width=0.45\textwidth]{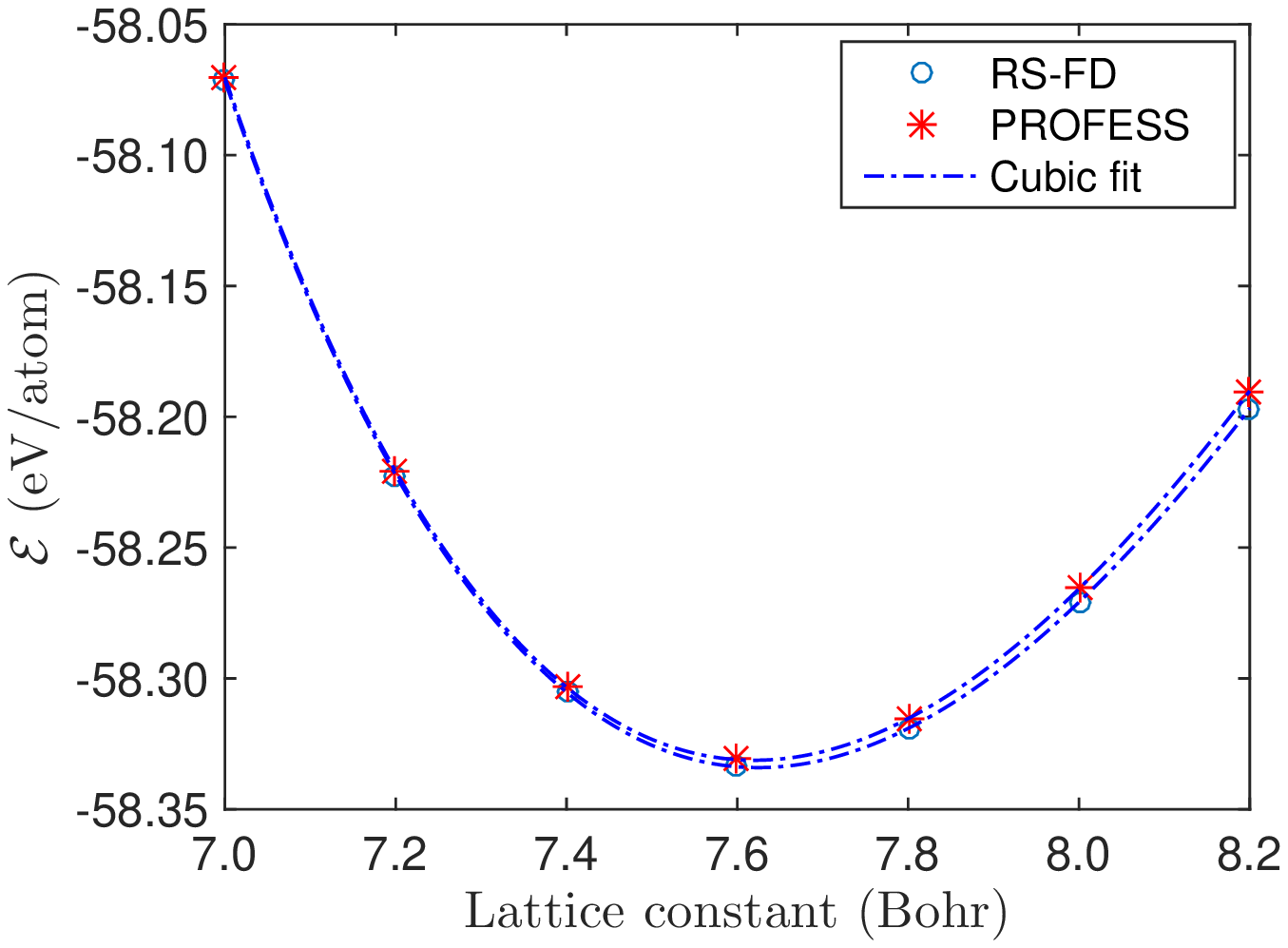}} \\
\caption{Variation of energy with lattice constant for FCC Aluminum.}
\label{Fig:EnergyLattice}
\end{figure}

Next, we determine the bulk properties of Aluminum for the TFW and WGC kinetic energy functionals using a supercell consisting of $5 \times 5 \times 5$ FCC unit cells ($M_a=500$). We start by calculating the energy/atom $\mathcal{E}$ using RS-FD and PROFESS for various lattice constants $a$, the results of which are presented in Fig. \ref{Fig:EnergyLattice}. We then employ a cubic spline fit to the data to determine the equilibrium lattice constant $a_e$ and the bulk modulus \cite{Finnis2003}
\begin{equation}
B = \frac{4}{9a_e} \frac{\partial^2 \mathcal{E}}{\partial a^2} \bigg|_{a_e} \,.
\end{equation}
In the above expression, $\mathcal{E}$ represents the energy of the primitive unit cell. It is evident from the results in Table \ref{Table:AluminumBulk} that the predictions of RS-FD are in very good agreement with PROFESS. In fact, the equilibrium lattice constants are identical to within $0.01$ Bohr for both the TFW and WGC functionals. The difference in the energy for the TFW and WGC functionals is $0.005$ eV/atom and $0.003$ eV/atom, respectively, with the difference in the bulk modulus being $0.006$ GPa and $0.859$ GPa, respectively. The slight difference in bulk modulus for the WGC functional can be attributed to the fact that RS-FD approximates the kernels $K_{mn}(|\bx-\bx'|)$ in Fourier space using rational functions. Indeed, using a larger number of rational functions to approximate $K_{mn}(|\bx-\bx'|)$ further improves the agreement between RS-FD and PROFESS. Specifically, for $R=6$, the difference in bulk modulus between PROFESS and RS-FD is $0.006$ GPa.

\begin{table}[H]
\centering
\begin{tabular}{ccccc}
\hline 
Kinetic energy &    Method    & $\mathcal{E}$ &      $a_e$         &    $B$  \\
 functional    &              &   (eV/atom)   &      (Bohr)        &    (GPa)      \\ 
\hline
      TFW      &     RS-FD    & $-60.226$ & $8.00$ & $57.3$ \\
               &    PROFESS   & $-60.221$ & $8.00$ & $57.3$ \\
\hline
      WGC      &     RS-FD    & $-58.335$ & $7.62$ & $68.1$ \\
               &    PROFESS   & $-58.332$ & $7.62$ & $68.9$ \\
\hline

\end{tabular}
\caption{Bulk properties of FCC Aluminum.}
\label{Table:AluminumBulk}
\end{table}

\subsubsection{Vacancy formation energy in Aluminum} \label{Subsec:Monovacancy}
\begin{figure}[H]
\centering
\subfloat[TFW]{\label{fig:TFWVacancyRelaxed}\includegraphics[keepaspectratio=true,width=0.45\textwidth]{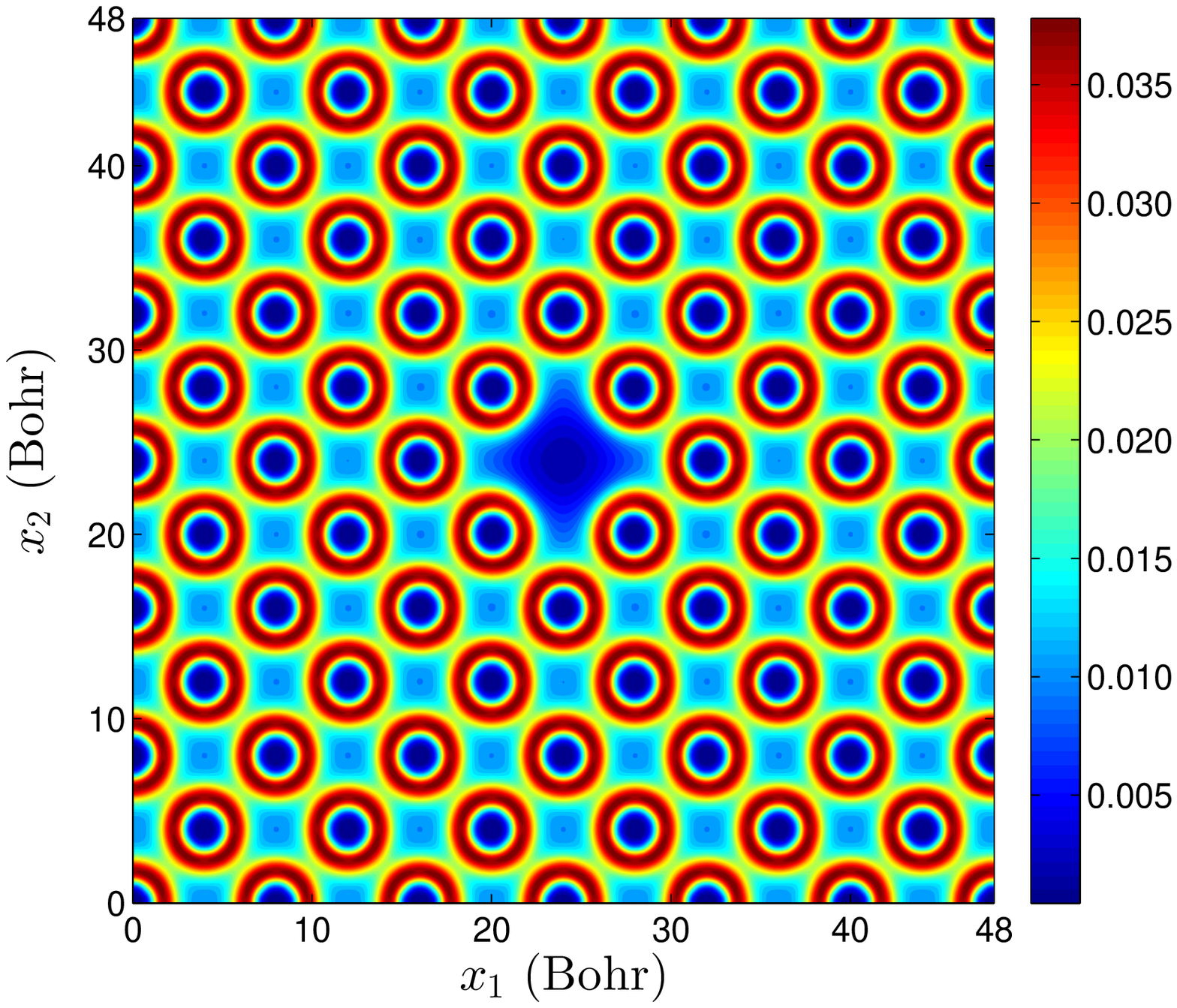}} 
\subfloat[WGC]{\label{fig:WGCVacancyRelaxed}\includegraphics[keepaspectratio=true,width=0.43\textwidth]{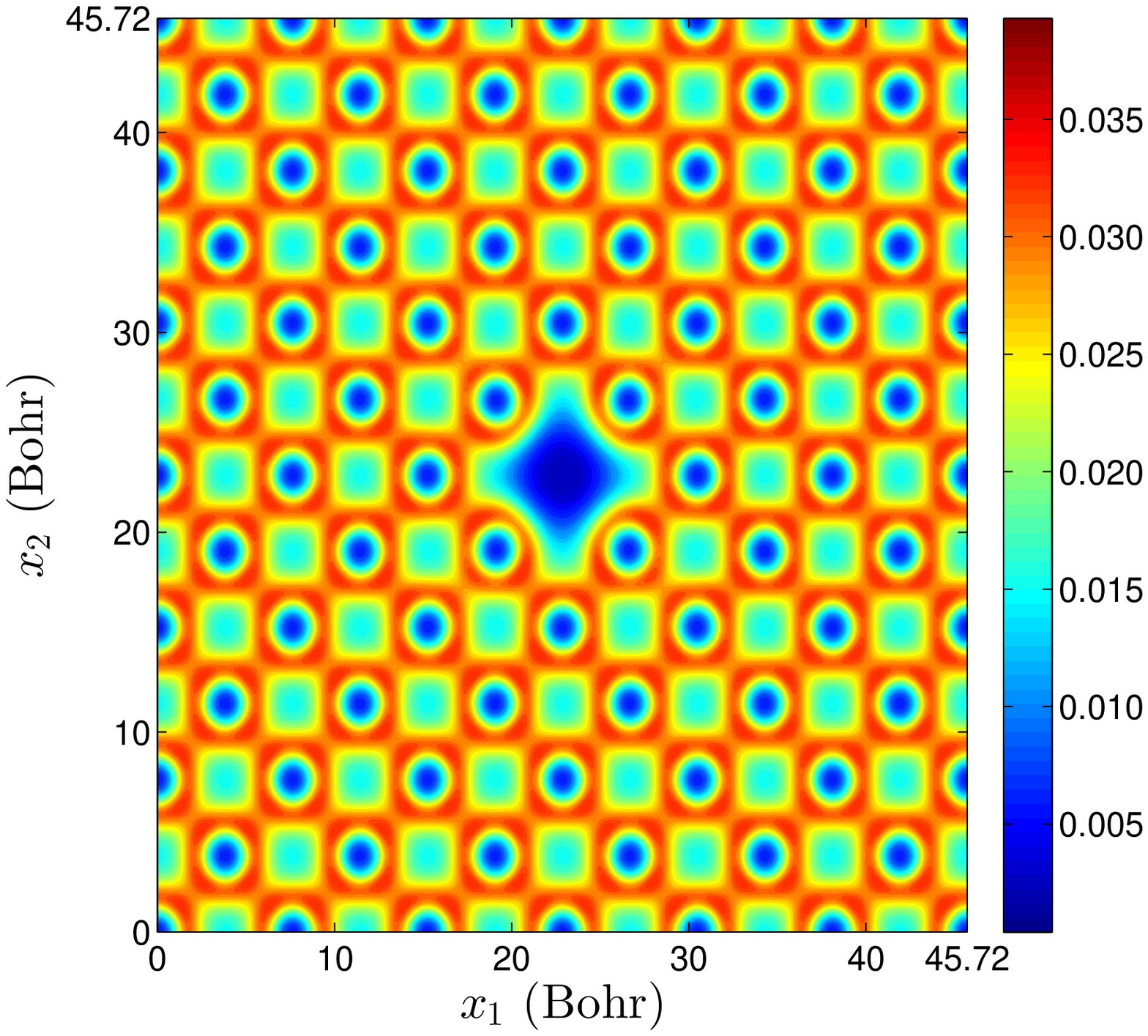}} 
\caption{Electron density contours on the mid-plane of a relaxed vacancy in stress-free FCC Aluminum.}
\label{Fig:ElecDensity}
\end{figure}

Finally, we calculate the vacancy formation energy in stress-free FCC Aluminum. We consider a supercell consisting of $6 \times 6 \times 6$ FCC Aluminum unit cells ($M_a=864$), and remove an atom from the center to create a vacancy. We calculate the vacancy formation energy $\mathcal{E}_{vf}$ using the relation \cite{Finnis2003,gillan1989calculation}
\begin{equation}
\mathcal{E}_{vf} =\mathcal{E}\left(M_a-1,1,\frac{M_a-1}{M_a} \Omega \right) - \left( \frac{M_a-1}{M_a} \right)\mathcal{E}(M_a,0,\Omega) \,,
\end{equation}
where $\mathcal{E}(M_a,n,\Omega)$ is used to denote the energy of a periodic cell $\Omega$ with $M_a$ occupied lattice sites and $n$ vacancies. We present the results so obtained in Table \ref{Table:VacancyFormationEnergy}, and plot the electron density contours on the mid-plane of a relaxed vacancy for the TFW and WGC kinetic energy functionals in Figs. \ref{fig:TFWVacancyRelaxed} and \ref{fig:WGCVacancyRelaxed}, respectively. We observe that the computed vacancy formation energies are in good agreement with PROFESS. In fact, the relaxed vacancy formation energies are identical to within $0.01$ eV and $0.02$ eV when using the TFW and WGC functionals, respectively. From the final relaxed configuration of the atoms, we find the maximum difference between the positions of the nuclei obtained by RS-FD and PROFESS to be $0.0025$ Bohr for the TFW functional and $0.016$ Bohr for the WGC functional. As discussed in the previous section, the larger discrepancy in WGC can be attributed to the approximate kernels $K_{mn}(|\bx-\bx'|)$ employed in RS-FD. 

\begin{table}[H]
\centering
\begin{tabular}{ccccc}
\hline 
Kinetic energy &    Method    & $a_e$  &  $\mathcal{E}_{vf}$ (unrelaxed) &  $\mathcal{E}_{vf}$ (relaxed)  \\
 functional    &              & (Bohr) &  (eV)  & (eV)    \\ 
\hline
      TFW      &     RS-FD    & $8.00$ & $0.87$  & $0.83$\\
               &    PROFESS   & $8.00$ & $0.87$  & $0.83$\\
\hline
      WGC      &     RS-FD    & $7.62$ & $0.60$  & $0.49$\\
               &    PROFESS   & $7.62$ & $0.59$  & $0.47$ \\
\hline
\end{tabular}
\caption{Vacancy formation energy in stress-free FCC Aluminum}
\label{Table:VacancyFormationEnergy}
\end{table}

\subsection{Scaling and Performance} \label{Subsec:ScalingPerformance}
We now study the scaling and performance of RS-FD for determining the electronic ground-state and atomic forces when linear response kinetic energy functionals are employed. In all the simulations, we utilize the WGC kinetic energy functional and a sixth-order finite-difference discretization with mesh size of $h=0.6927$ Bohr. We choose a cutoff radius of $8.0$ Bohr for the charge densities of the nuclei, whereby the enclosed charge for each nucleus is accurate to within $2 \times 10^{-6}$. In addition, we employ tolerances of $1 \times 10^{-3}$, $1 \times 10^{-6}$, and $1 \times 10^{-2}$ on the normalized residual for convergence of the conjugate gradient, GMRES, and fixed-point methods, respectively. The energies and forces so obtained are converged to within the chemical accuracy of $0.027$ eV/atom and $0.027$ eV/Bohr, respectively.

\begin{figure}[H]
\centering
\subfloat[Strong scaling]{\label{fig:StrongScaling}\includegraphics[keepaspectratio=true,width=0.45\textwidth]{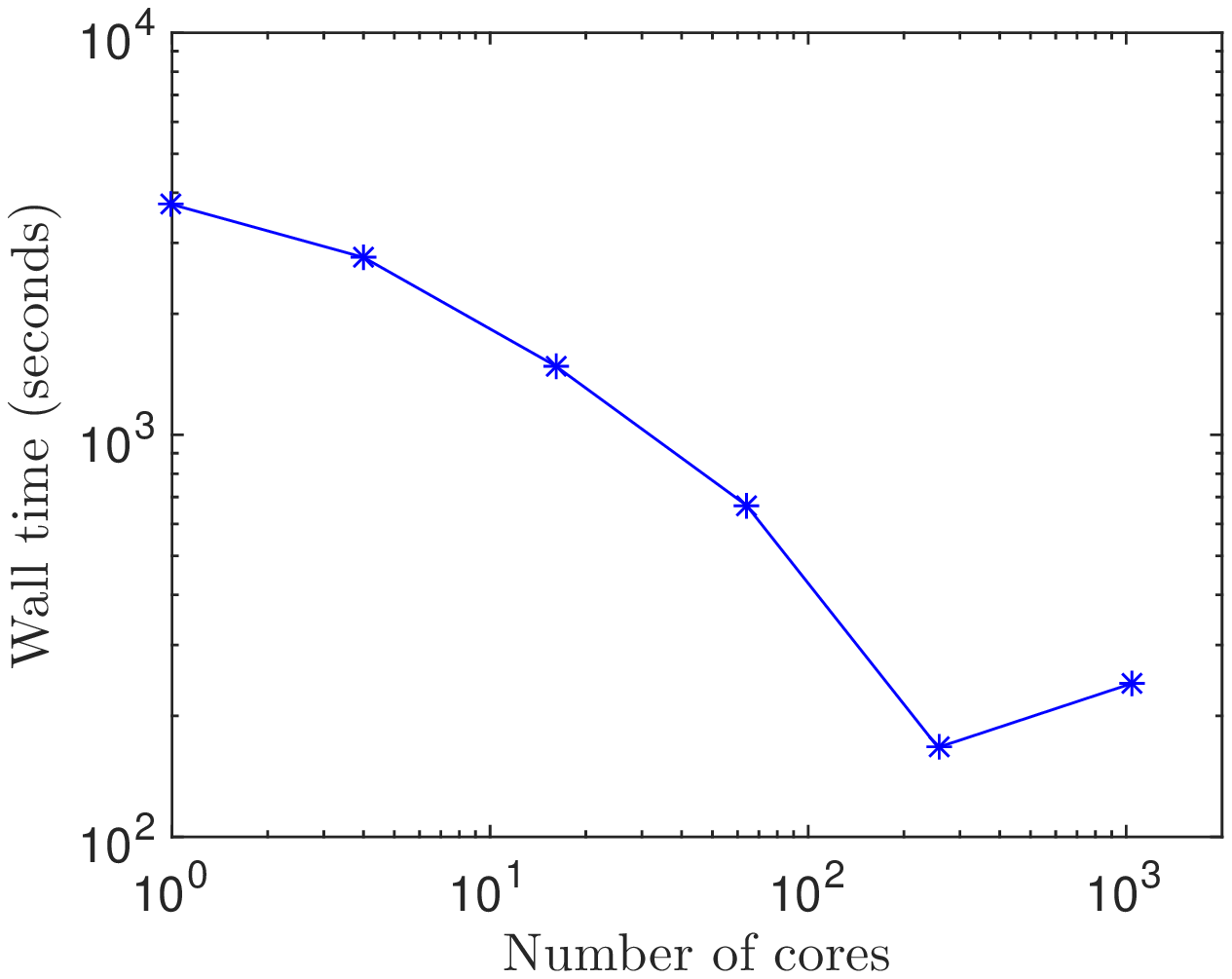}} 
\subfloat[Weak scaling]{\label{fig:WeakScaling}\includegraphics[keepaspectratio=true,width=0.45\textwidth]{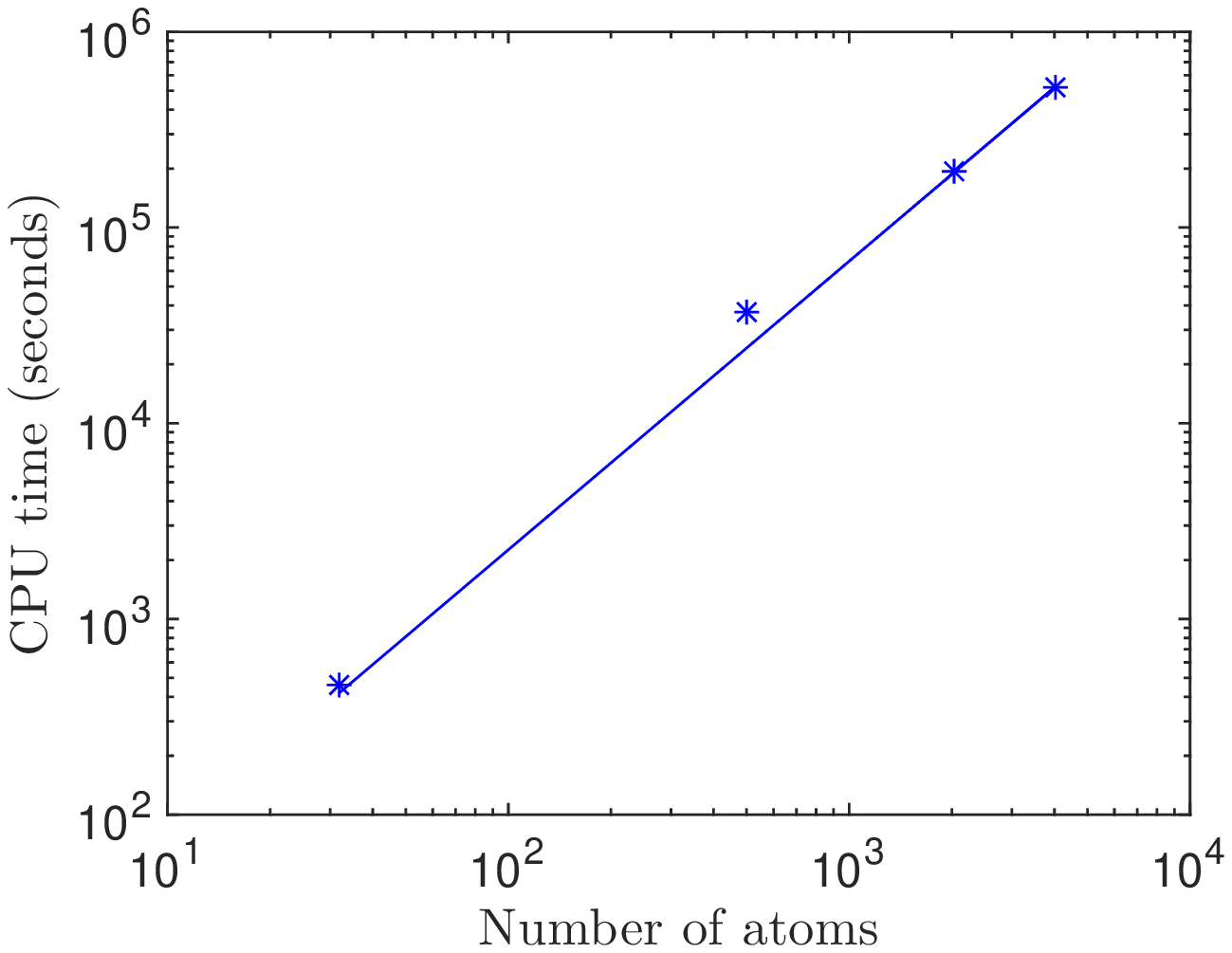}} 
\caption{Strong and weak scaling of RS-FD.}
\label{Fig:Scaling}
\end{figure}

We first analyze the strong scaling of RS-FD for the system consisting of $5\times 5 \times 5$ FCC unit cells of Aluminum (Al$_{500}$). In Fig. \ref{fig:StrongScaling}, we present the wall time for performing the simulation on $1$, $4$, $16$, $64$, $256$, and $1024$ computational cores. We observe that there is a steady decrease in the wall time up to $256$ cores, which then increases for $1024$ cores. Specifically, the wall time of $167$ seconds on $256$ cores represents a factor of $22.34$ reduction compared to $1$ core, and a factor of $8.90$ reduction compared to $16$ cores. The sudden increase in wall time for $1024$ cores is due to the finite-difference order becoming larger than the number of finite-difference nodes local to each core in each direction. Next, we study the weak scaling of RS-FD by determining the CPU time for (i) $2\times 2 \times 2$ FCC unit cells of Aluminum (Al$_{32}$) on $4$ cores (ii) $5\times 5 \times 5$ FCC unit cells of Aluminum (Al$_{500}$) on $64$ cores (iii) $8\times 8 \times 8$ FCC unit cells of Aluminum (Al$_{2048}$) on $256$ cores (iv) $10\times 10 \times 10$ FCC unit cells of Aluminum (Al$_{4000}$) on $512$ cores. The systems and number of cores have been chosen such that there are approximately $8$ atoms per core. We present the results so obtained in Fig. \ref{fig:WeakScaling}. From a curve fit to the data, we find that RS-FD has an overall scaling of $\mathcal{O}(M_a^{1.47})$ with respect to the number of atoms. 
 
The practical scaling of RS-FD being worse than $\mathcal{O}(M_a)$ merits further consideration. The number of iterations required by the fixed-point method---$3$ for all the examples considered here---is independent of system size, a feature necessary for achieving $\mathcal{O}(M_a)$ scaling. However, the number of iterations required by the conjugate gradient method (including inner and outer iterations) increases from $667$ to $1391$ as the system size increases from Al$_{32}$ to Al$_{4000}$. The corresponding total number of GMRES iterations for the Helmholtz (all $52$ linear systems) and Poisson equations increases from $79,295$ to $91,510$ and $2252$ to $44,166$, respectively. This increase in iterations with number of atoms is the underlying reason why RS-FD is unable to achieve $\mathcal{O}(M_a)$ scaling in practical computations. This motivates the development of real-space preconditioners for the conjugate gradient method (Algorithm \ref{Algo:NLCG}) and the use of more sophisticated preconditioning schemes like multigrid \cite{hackbusch2013multi} for the GMRES method. Specifically, effective preconditioning techniques that render the number of iterations independent of system size will enable RS-FD to achieve $\mathcal{O}(M_a)$ scaling in practice.

Finally, we comment on the relative performance of RS-FD and the plane-wave code PROFESS. PROFESS demonstrates tremendously superior CPU times when compared to RS-FD, mainly because its ability to solve the Poisson equation---arising for every update in the square-root electron density---using the Fast Fourier Transform (FFT). In addition, unlike real-space methods where the Helmholtz equations need be solved for determining the kernel potential, plane-wave approaches can efficiently evaluate the convolutions directly using FFT. However, it is highly challenging to efficiently scale FFT's to modern large-scale distributed memory computer architectures that regularly contain thousands of cores. Therefore, RS-FD is expected to become competitive with plane-wave codes like PROFESS in the context of wall times for large-scale parallel simulations, particularly when effective preconditioning schemes are employed. For example, using a maximum of the available $1024$ cores, the least wall time achieved by RS-FD for the Al$_{4000}$ system is $508$ seconds ($1024$ cores), which is only about two times larger than the smallest wall time achieved by PROFESS ($32$ cores). Overall, RS-FD represents an attractive approach for performing large-scale parallel simulations consisting of thousands of atoms. 

\section{Concluding Remarks} \label{Section:Conclusions}
We have presented a real-space formulation and higher-order finite-difference implementation of periodic Orbital-free Density Functional Theory (OF-DFT). Specifically, utilizing a local reformulation of the electrostatic and kernel energy/potential terms, we have developed a generalized framework for performing OF-DFT simulations that is able to accommodate different variants of the electronic kinetic energy. In particular, for linear-response kinetic energy functionals, we have proposed a fixed-point technique that is similar in spirit to the self-consistent field (SCF) method employed in DFT calculations. We have also developed a parallel finite-difference implementation of this formulation, using which we have demonstrated that higher-order finite-differences are necessary to efficiently obtain chemical accuracies in the energy and forces. Additionally, we have established that the fixed-point iteration accelerated using Anderson mixing converges rapidly in about $5$ iterations. We have validated the accuracy of our results by comparing the energies and forces with plane-wave methods for selected examples. Overall, we conclude that the suitability of the proposed formulation and implementation for scalable high performance computing make them an attractive choice for performing large-scale OF-DFT calculations consisting of thousands of atoms.

We finally note that higher-order finite-differences are an appealing discretization scheme for electronic structure calculations based on OF-DFT. This is due to their simplicity, potential for scalability to massively-parallel distributed-memory computer architectures, and ability to achieve chemical accuracies desired in electronic structure calculations. The authors are currently pursuing effective real-space preconditioners that will enable the RS-FD implementation to scale linearly with respect to the number of atoms. Finally, extending the current work to finite-temperatures and therefore enabling molecular dynamics simulations is a worthy subject of future research. 

\section*{Acknowledgements}
The authors gratefully acknowledge the support of Office of Naval Research (N00014-13-C-0267), the start-up funds provided by Georgia Institute of Technology and the National Science Foundation (Award number: 1333500) 

\appendix 
\section{Coefficients in the Helmholtz equations for the WGC kinetic energy functional} \label{Appendix:WGC}
The coefficients $P_{mnr}$ and $Q_{mnr}$ arising in the Helmholtz equations (Eqns. \ref{Eqn:Helmholtz:beta} and \ref{Eqn:Helmholtz:alpha})  are determined by fitting the kernels $K_{mn}(|\bx-\bx'|)$ in Fourier space using rational functions \cite{Choly2002}. In Table \ref{Table:ConstantsHelmholtz}, we present the values for the WGC functional when a fourth-order expansion ($R=4$) is employed. The coefficients satisfy the relations $P_{mnr}=P_{nmr}$ and $Q_{mnr}=Q_{nmr}$, with $P_{mn2}=P_{mn1}^*$, $Q_{mn2}=Q^*_{mn1}$, $P_{mn4}=P_{mn3}^*$ and $Q_{mn4}=Q^*_{mn3}$. Here, the superscript $*$ denotes the complex conjugate. 

\begin{table}[H]
\centering
\begin{tabular}{ccc}
\hline 
Coefficients        & $r=1$                    &  $r=3$  \\
\hline
$P_{00r}$       & $+0.108403+i0.079657$   &  $-0.908403+i0.439708$   \\
$Q_{00r}$       & $-0.470923-i0.465392$   &  $+0.066051-i0.259678$     \\
$P_{10r}$       & $-0.030515+i0.015027$   &  $+0.028915-i0.008817$     \\
$Q_{10r}$       & $-0.597793-i0.294130$   &  $-0.087917-i0.164937$     \\
$P_{20r}$       & $+0.008907-i0.032841$   &  $-0.034974+i0.009116$   \\
$Q_{20r}$       & $-0.537986-i0.233840$   &  $-0.041565-i0.196662$   \\
$P_{11r}$       & $+0.012423-i0.034421$   &  $-0.031907+i0.007392$   \\
$Q_{11r}$       & $-0.511699-i0.0266195$  &  $-0.034031-i0.188927$   \\
\hline
\end{tabular}
\caption{Coefficients in the Helmholtz equations (Eqns. \ref{Eqn:Helmholtz:beta} and \ref{Eqn:Helmholtz:alpha}) for the WGC kinetic energy functional \cite{Choly2002}.}
\label{Table:ConstantsHelmholtz}
\end{table}   

\section{Electrostatic correction for overlapping charge density of nuclei} \label{Appendix:Correct:RepulsiveEnergy}
In the local reformulation of the electrostatics presented in Section \ref{Subsec:ElectrostaticReformulation}, the repulsive energy can be expressed as 
\begin{equation} \label{Eqn:EZZ:rf}
\mathcal{E}_{\rm zz}(\bR) = \frac{1}{2} \int_{\Omega} \int_{\R^3} \frac{b(\bx,\bR) b(\bx',\bR)}{|\bx-\bx'|} \, \mathrm{d\bx'} \, \mathrm{d\bx} - \frac{1}{2}\sum_{J} \int_{\Omega} b_J(\bx,\bR_J) V_J(\bx,\bR_J) \, \mathrm{d\bx} \,,
\end{equation}
where the second term accounts for the self energy of the nuclei. Using Eqn. \ref{Eqn:b:Pseudopotential}, we arrive at
\begin{eqnarray}
\mathcal{E}_{\rm zz}(\bR) & = & \frac{1}{2} \sum_{I} \sum_{J} \int_{\Omega} b_I(\bx,\bR_I) V_J(\bx,\bR_J) \, \mathrm{d\bx} - \frac{1}{2}\sum_{J} \int_{\Omega} b_J(\bx,\bR_J) V_J(\bx,\bR_J) \, \mathrm{d\bx} \nonumber \\
& = & \frac{1}{2} \sum_{I} \sum_{ \begin{subarray}{c} J \\ J \neq I \end{subarray}} \int_{\Omega} b_I(\bx,\bR_I) V_J(\bx,\bR_J) \, \mathrm{d\bx} \label{Eqn:b:Ezz} \,.
\end{eqnarray}
Above, the summations indices $I$ and $J$ run over all atoms in $\R^3$. If the charge density of the nuclei do not overlap, Eqn. \ref{Eqn:b:Ezz} can be rewritten as 
\begin{eqnarray}
\mathcal{E}_{\rm zz}(\bR) & = &  \frac{1}{2} \sum_{I} \sum_{ \begin{subarray}{c} J \\ J \neq I \end{subarray}} Z_J \int_{\Omega} \frac{b_I(\bx,\bR_I)}{|\bx-\bR_J|} \, \mathrm{d\bx}  = \frac{1}{2} \sum_{I} \sum_{ \begin{subarray}{c} J_{\Omega} \\ J_{\Omega} \neq I \end{subarray}} Z_{J_{\Omega}} V_I(\bR_{J_\Omega},\bR_I)\nonumber \\
& = & \frac{1}{2} \sum_{I} \sum_{\begin{subarray}{c} J_{\Omega} \\J_{\Omega} \neq I \end{subarray}} \frac{Z_{I} Z_{J_{\Omega}}}{|\bR_{I}-\bR_{J_{\Omega}}|} \label{Eqn:RepulsiveEnergyAppendix} \,,
\end{eqnarray}
which is exactly the expression given in Eqn. \ref{Eqn:EZZ} for the repulsive energy prior to reformulation. However, the use of relatively `soft' pseudopotentials --- which are attractive because of the significant reduction in the number of basis functions required for convergence --- can frequently result in overlapping charge density of the nuclei. Even in this situation, the repulsive energy in ab-initio calculations is calculated by treating the nuclei as point charges (i.e., Eqn. \ref{Eqn:RepulsiveEnergyAppendix}). Since the electrostatic reformulation in this work does make this distinction between overlapping and non-overlapping charge density of the nuclei, we present a technique below that reestablishes agrement. 

We start by generating a `reference' charge density 
\begin{equation}
\tilde{b}(\bx,\bR) = \sum_{J} \tilde{b}_J(\bx,\bR_J) \,,
\end{equation}
which is the superposition of spherically symmetric and compactly supported `reference' charge densities $\tilde{b}_J (\bx,\bR_J)$. These nuclei-centered charge densities satisfy the relations
\begin{equation}
\int_{\R^3} \tilde{b}_J (\bx,\bR_J) \, \mathrm{d\bx} = Z_J \,, \quad \int_{\Omega} \tilde{b}(\bx,\bR) \mathrm{d\bx} = N_e \,.
\end{equation}
Thereafter, the correction to the repulsive energy can be expressed as
\begin{eqnarray} \label{Eqn:RepulsiveCorrection}
\mathcal{E}_c^*(\bR) & = & \frac{1}{2} \int_{\Omega} \int_{\R^3} \frac{\tilde{b}(\bx,\bR) \tilde{b}(\bx',\bR)}{|\bx-\bx'|} \, \mathrm{d\bx'} \, \mathrm{d\bx} - \frac{1}{2} \int_{\Omega} \int_{\R^3} \frac{b(\bx,\bR) b(\bx',\bR)}{|\bx-\bx'|} \, \mathrm{d\bx'} \, \mathrm{d\bx}  \nonumber \\
& & - \frac{1}{2}\sum_{J} \int_{\Omega} \tilde{b}_J(\bx,\bR_J) \tilde{V}_J(\bx,\bR_J) \, \mathrm{d\bx}  + \frac{1}{2}\sum_{J} \int_{\Omega} b_J(\bx,\bR_J) V_J(\bx,\bR_J) \, \mathrm{d\bx} .
\end{eqnarray}
A direct computation of this energy correction will scale quadratically with respect to the number of atoms. In order to enable linear-scaling, we rewrite Eqn. \ref{Eqn:RepulsiveCorrection} as
\begin{eqnarray} \label{Eqn:RepulsiveCorrection2}
\mathcal{E}_c^*(\bR) & = & \frac{1}{2} \int_{\Omega} \left( \tilde{b}(\bx,\bR) + b(\bx,\bR) \right) V_c(\bx,\bR) \, \mathrm{d\bx} + \frac{1}{2}\sum_{J} \int_{\Omega} b_J(\bx,\bR_J) V_J(\bx,\bR_J) \, \mathrm{d\bx} \nonumber \\
& & - \frac{1}{2}\sum_{J} \int_{\Omega} \tilde{b}_J(\bx,\bR_J) \tilde{V}_J(\bx,\bR_J) \, \mathrm{d\bx} \,,
\end{eqnarray}
where $V_c(\bx,\bR)$ is the solution to the Poisson equation 
\begin{equation}
\frac{-1}{4\pi} \nabla^2 V_c(\bx,\bR) = \tilde{b}(\bx,\bR) - b(\bx,\bR)
\end{equation} 
with periodic boundary conditions. The potential $V_c(\bx,\bR)$ so calculated is accurate to within a constant, which can be determined by evaluating $\sum_{J}(V_J(\bx,\bR_J) - \tilde{V}_J(\bx,\bR_J))$ at any point in space. 

The correction to the forces on the nuclei 
\begin{equation}
{\bf f}_J^c = -\frac{\partial \mathcal{E}_c^*(\bR)}{\partial \bR_J} 
\end{equation}
can be represented as 
\begin{eqnarray}
{\bf f}_J^c & = & -\frac{1}{2} \sum_{J'} \int_{\Omega} \bigg[ \frac{\partial \tilde{b}_{J'}(\bx,\bR_{J'})}{\partial \bR_{J'}}\left(V_c(\bx,\bR)- \tilde{V}_{J'}(\bx,\bR_{J'})\right) + \frac{\partial b_{J'}(\bx,\bR_{J'})}{\partial \bR_{J'}}\left(V_c(\bx,\bR)+V_{J'}(\bx,\bR_{J'})\right) \nonumber \\ 
& & + \frac{\partial V_c(\bx,\bR)}{\partial \bR_{J'}}\left(\tilde{b}(\bx,\bR)+b(\bx,\bR)\right) + b_{J'}(\bx,\bR_{J'})\frac{\partial V_{J'}(\bx,\bR_{J'})}{\partial \bR_{J'}} - \tilde{b}_{J'}(\bx,\bR_{J'})\frac{\partial \tilde{V}_{J'}(\bx,\bR_{J'})}{\partial \bR_{J'}} \bigg] \,\mathrm{d\bx}  \nonumber \\
& = & \frac{1}{2} \sum_{J'} \int_{\Omega} \bigg[ \nabla \tilde{b}_{J'}(\bx,\bR_{J'}) \left(V_c(\bx,\bR)- \tilde{V}_{J'}(\bx,\bR_{J'})\right) + \nabla b_{J'}(\bx,\bR_{J'}) \left(V_c(\bx,\bR)+V_{J'}(\bx,\bR_{J'})\right) \nonumber \\ 
& & + \nabla V_{c,J'}(\bx,\bR_{J'}) \left(\tilde{b}(\bx,\bR)+b(\bx,\bR)\right) + b_{J'}(\bx,\bR_{J'}) \nabla V_{J'}(\bx,\bR_{J'}) - \tilde{b}_{J'}(\bx,\bR_{J'}) \nabla \tilde{V}_{J'}(\bx,\bR_{J'}) \bigg] \,\mathrm{d\bx} \nonumber \,,
\end{eqnarray}
where the summation $J'$ is over $J^{th}$ atom and its periodic images. Additionally, 
\begin{equation}
\nabla V_{c,J'}(\bx,\bR_{J'}) = \nabla \tilde{V}_{J'}(\bx,\bR_{J'}) - \nabla V_{J'}(\bx,\bR_{J'}) .
\end{equation}
It is important to note that even with these corrections to the energy and forces, the overall OF-DFT formulation maintains its linear-scaling nature with respect to the number of atoms. 


\section{Conjugate gradient method for OF-DFT}\label{Appendix:NLCGTeter}
In Algorithm \ref{Algo:NLCG}, we present the conjugate gradient method implemented in RS-FD to solve the variational problem in Eqn. \ref{Eqn:MVF}. This differs from the standard non-linear conjugate gradient method \cite{Shewchuk1994} in that it is able to handle the constraints $\mathcal{C}(u)=0$ and $u \geq 0$. 

\begin{algorithm}[H] \label{Algo:NLCG}
{\bf Input}: $u_0$, $\bR$, $V_{LR}$ and $N_{restart}$  \\
$q=0$ \\
\Repeat(){$\norm{r} < tol$}
{
$\eta_q = \frac{1}{N_e} \langle u_q,\mathcal{H} u_q \rangle $, where $\langle .,. \rangle$ denotes the inner product \\
$r_q = -2(\mathcal{H}u_q - \eta_q u_q)$ \\
$\xi = \frac{\langle r_q-r_{q-1} ,r_q\rangle}{\langle r_{q-1},r_{q-1} \rangle}$ \\
\eIf{$q= m N_{restart}$ ($m\in \mathbb{N}$) or $\xi \leq 0$}
{$d_q=r_q$}
{$d_q = r_q + \xi d_{q-1}$}
$s = \arg \inf_{s\in \R} \mathcal{\hat{E}}\left(\sqrt{N_e} \frac{u_q-sr_q}{\norm{u_q-sr_q}},\bR,V_{LR}\right)$ \\
$u_{q+1} = \sqrt{N_e} \frac{u_q-sr_q}{\norm{u_q-sr_q}}$\\
$q = q + 1$\\
}
{\bf Output}:\, $u = u_{q}$ 
\caption{Non-linear conjugate gradient method for OF-DFT}
\end{algorithm} 


\section{Anderson Mixing}\label{Appendix:Anderson}
The fixed-point problem in Eqn. \ref{Eqn:FixedPoint:Map1} can be rewritten as 
\begin{equation}
f(V_{LR}) = 0 \,, \quad f(V_{LR}) = \mathcal{V} \big[ \mathcal{U}(V_{LR}) \big] - V_{LR} \,.
\end{equation}
This equation can be solved using an iteration of the form \cite{fang2009two,lin2013elliptic}
\begin{equation}
V_{LR,k+1} = V_{LR,k} - C_k f(V_{LR,k})  \,,
\end{equation}
where $C_k$ is chosen to approximate the inverse Jacobian. In multi-secant type methods, $C_k$ is set to the solution of the constrained minimization problem \cite{fang2009two,lin2013elliptic}
\begin{equation} \label{Eqn:Ck_update}
\min_{C} \frac{1}{2} \norm{C-C_{k-1}}_2^2 \quad \text{s.t.} \quad S_k = C Y_k \,,  
\end{equation}
where
\begin{eqnarray}
S_k & = &[V_{LR,k-m+1} - V_{LR,k-m}, \ldots ,V_{LR,k} - V_{LR,k-1}] \,,  \nonumber \\
Y_k & = &[f(V_{LR,k-m+1}) - f(V_{LR,k-m}), \ldots ,f(V_{LR,k}) - f(V_{LR,k-1})] \,.  \nonumber 
\end{eqnarray}
In the above equations, $m$ represents the mixing history. The solution of this variational problem is
\begin{equation}
C_k = C_{k-1} + (S_k-C_{k-1}Y_k)(Y_k^T Y_k)^{-1}Y_k^T \,.
\end{equation}
In the specific case of Anderson mixing \cite{anderson1965iterative}, $C_{k-1}$ is set to $-\zeta I$, where $I$ is a $m\times m$ identity matrix. This leads to the update formula:
\begin{equation}
V_{LR,k+1} = V_{LR,k} + \zeta f(V_{LR,k}) - (S_k+\zeta Y_k) (Y_k^T Y_k)^{-1} Y_k^T f(V_{LR,k}) \,.
\end{equation}

\bibliographystyle{ReferenceStyle}

\end{document}